\documentclass[
 reprint,
superscriptaddress,
nofootinbib,
 amsmath,amssymb,
 aps,
 prd,
]{revtex4-2}

\usepackage{graphicx}
\usepackage{dcolumn}
\usepackage{bm}
\usepackage{hyperref}
\usepackage[mathlines]{lineno}
\usepackage{enumitem} 
\usepackage{duckuments}

\usepackage[normalem]{ulem}

\usepackage{color}

\newcommand{\aref}[1]{\hyperref[#1]{Appendix~\ref*{#1}}}

\begin{document}

\preprint{APS/123-QED}

\title{Finite Populations \& Finite Time: The Non-Gaussianity of a Gravitational Wave Background}

\author{William G. Lamb}
\affiliation{Department of Physics and Astronomy, Vanderbilt University, 2301 Vanderbilt Place, Nashville, TN 37235, USA}

\author{Jeremy M. Wachter}
\affiliation{School of Sciences \& Humanities, Wentworth Institute of Technology, Boston, MA 02115}

\author{Andrea Mitridate}
\affiliation{Deutsches Elektronen-Synchrotron DESY, Notkestr. 85, 22607 Hamburg, Germany}

\author{Shashwat C. Sardesai}
\affiliation{University of Wisconsin Milwaukee, Milwaukee, WI 53211}

\author{Bence B\'ecsy}
\affiliation{Institute for Gravitational Wave Astronomy and School of Physics and Astronomy, University of Birmingham, Edgbaston, Birmingham B15 2TT, UK}

\author{Emily L. Hagen}
\affiliation{School of Sciences \& Humanities, Wentworth Institute of Technology, Boston, MA 02115}

\author{Stephen R. Taylor}
\affiliation{Department of Physics and Astronomy, Vanderbilt University, 2301 Vanderbilt Place, Nashville, TN 37235, USA}

\author{Luke Zoltan Kelley}
\affiliation{Department of Astronomy, University of California, Berkeley, 501 Campbell Hall \#3411, Berkeley, CA 94720, USA}

\date{\today}

\begin{abstract}
Strong evidence for an isotropic, Gaussian gravitational wave background (GWB) has been found by multiple pulsar timing arrays (PTAs). The GWB is expected to be sourced by a finite population of supermassive black hole binaries (SMBHBs) emitting in the PTA sensitivity band, and astrophysical inference of PTA data sets suggests a GWB signal that is at the higher end of GWB spectral amplitude estimates. However, current inference analyses make simplifying assumptions, such as modeling the GWB as Gaussian, assuming that all SMBHBs only emit at frequencies that are integer multiples of the total observing time, and ignoring the interference between the signals of different SMBHBs. In this paper, we build analytical and numerical models of an astrophysical GWB from circular, inspiralling binaries inclined relative to the line-of-sight of the observer, without the above approximations, and compare the statistical properties of its induced PTA signal to those of a signal produced by a Gaussian GWB. We show that finite population and windowing effects introduce non-Gaussianities in the PTA signal, which are currently unmodeled in PTA analyses.
\end{abstract}

\maketitle

\section{Introduction}

Multiple pulsar timing arrays (PTAs) have found strong evidence for a nanohertz-frequency gravitational-wave background (GWB) \citep{NANOGrav:2023gor, Antoniadis:2023rey, Reardon:2023gzh, Xu:2023wog, miles2025meerkat}, with significant effort being coordinated by individual PTAs and collaboratively as the International Pulsar Timing Array \citep{manchester2013international} to determine its source(s). However, analyses to determine its (dominant) source have so far been inconclusive \citep{ng15_astro, NANOGrav:2023hvm, agazie2023nanograv_aniso, agazie2023cw, antoniadis2024second_4, antoniadis2024second_5, grunthal2025meerkat, agazie2024nanograv_discrete, ellis2023prospects}.

The two families of sources that could produce a GWB at nanohertz frequencies are cosmological and astrophysical. The cosmological description results from early universe phenomena (see, for example, Refs.~\citep{caprini2018cosmological, NANOGrav:2023hvm}), and a detection would result in the first direct probe of the pre-recombination universe. Alternatively, the astrophysical description identifies the source as a finite population of supermassive black hole binaries (SMBHBs) inspiralling from gravitational wave (GW) radiation \citep{Rajagopal_1995, jaffe2003gravitational, wyithe2003low, sesana2004low, burke2019astrophysics}. The GWs emitted by unresolved binaries contribute to the GWB, while a number of resolved sources will result in continuous GW signals that are louder than the nanohertz GWB and may be resolvable \citep{sesana2009gravitational, rosado2015expected, kelley2018single, burke2019astrophysics, becsy2022exploring}, but have yet to be detected \citep{chen2026searching, agazie2023cw, agarwal2026nanograv, antoniadis2024second_5, IPTA:2023ero, zhao2025searching, zhao2026targeted}.

We can characterize the GWB by expanding the metric perturbation it generates in plane waves:
\begin{equation} 
    h_{ab}(t,\vec{x})=\sum_A\int df\int d\hat\Omega\, \tilde h_A(f,\hat\Omega) e^{i 2\pi f(t-\hat\Omega\cdot\vec{x})}\epsilon_{ab}^A(\hat{\Omega})\,,
\end{equation}
where $f$ is the GW frequency, $\hat\Omega$ the direction of propagation of the plane waves, $A=+,\times$ labels the two GW polarizations, $\epsilon_{ab}^A$ are the GW polarization tensors, and $\tilde h_A(f,\hat\Omega)$ are two complex functions (one for each GW polarization) satisfying $\tilde h_A^*(f,\hat\Omega)=\tilde h_A(-f,\hat\Omega)$. A GWB is characterized by the fact that the functions $\tilde h_A(f,\hat \Omega)$ can be treated as random variables, whose distribution is set by the properties of the GWB source. Typically, several assumptions are made when describing the statistics of the $\tilde h_A^*(f,\hat\Omega)$ functions:
\begin{enumerate}[label=(\roman*)]
  \item Gaussianity---since the GWB is expected to arise as a central-limit-theorem process, it is common to model it as a Gaussian process. Specifically, the $\tilde{h}_A^*(f,\hat\Omega)$ functions are assumed to be zero-mean Gaussian random variables.
  \item Isotropy---the statistical properties of $\tilde h_A^*(f,\hat\Omega)$ are assumed to be the same in all sky directions.
  \item Unpolarized---the statistical properties of both GW polarizations are assumed to be the same.
  \item Stationary---the properties of the GWB are assumed to be time-independent.
\end{enumerate}
With all these assumptions, the two-point functions for the $\tilde h_A^*(f,\hat\Omega)$ functions reads:
\begin{equation}\label{eq:gaussian_gwb_corr}
        \left<\tilde{h}_A^*(f, \hat\Omega) \tilde{h}_{A'}(f', \hat\Omega')\right> = \delta_{AA'}\delta(f- f') \delta^2(\hat{\Omega}, \hat{\Omega}') \frac{S_h(f)}{16\pi},
\end{equation}
where $S_h(f)$ is the GW strain power spectral density (PSD).
These assumptions are valid for both cosmologically sourced GWBs \citep{leshouches, caprini2018cosmological} and for an astrophysical GWB if there is a large number of unresolvable sources \citep{allen2023variance, allen_valtolina_24}.
However, for a low number of sources, GW shot noise will drive non-Gaussianity \citep{allen2023variance, allen_valtolina_24, lamb_taylor_24, sato2024exploring, cornish2015gravitational}.
GW shot noise will also result in anisotropies in addition to clustering of sources by large-scale structure  \citep{sato2024exploring, allen2024source}, while inclined binaries will generate polarization in the GWB \citep{sato2024exploring}. Neither anisotropy \citep{chen2026searching, agazie2023nanograv_aniso} nor polarization \citep{jimenez2024measuring} has been detected by PTAs yet.

Comparison of current GWB spectrum measurements with predictions from SMBHB population models suggests a possible underestimate of the number density or the masses of sources \citep{ng15_astro, antoniadis2024second_4, sato2025distribution, sato2024supermassive, toubiana2025reconciling}, or mismodeled noise in PTA analyses \citep{crisostomi2025beyond, di2024systematic, di2025choosing}.
Model selection studies prefer a Gaussian cosmological source for the GWB over the fiducial Gaussian power-law model in the NANOGrav 15yr data set \citep{NANOGrav:2023hvm}; however, as we will show in this paper, the Gaussian power-law does not properly model the astrophysical GWB, which hinders any conclusions regarding the source of the GWB.

Deviations from the power-law spectrum are driven by Poisson variance of the finite population \citep{Sesana:2008mz, becsy2022exploring, lamb_taylor_24}, and mismodeling the GWB can result in biased parameter estimation.
For example, if there are resolvable continuous waves above a Gaussian power-law GWB, failure to model the continuous wave will result in over-estimation of the power-law amplitude and under-estimation of its spectral index \citep{becsy2023detect}. 

Instead, we can model a superposition of continuous sources above the power-law \citep{becsy2020joint, becsy2023detect}.
Unfortunately, this is difficult to analyze with real data because of its slow likelihood evaluation time, unknown number of resolvable continuous wave sources, and very large parameter space.
Instead, simpler spectral models that are computationally tractable are often used.
For example, Poisson variance can be modeled as `excursions' from a power-law, either by using data-driven weights to model deviations \citep{NANOGrav:2023gor, sardesai2024characterizing} or by making no assumptions about the shape of the spectrum \citep{PhysRevD.87.104021, agazie2024nanograv_discrete}.
These spectra can be useful to point towards any continuous wave sources; however, the intrinsic Gaussianity of these spectra will result in mismodeling and biased parameter estimation \citep{Xue:2024qtx}.

More involved astrophysical inference techniques compare ensemble distributions of GWB spectra, obtained through density estimation of many ensemble realizations of semi-analytic SMBHB population models, to posteriors from Bayesian analyses such as Markov Chain Monte Carlo (MCMC) \cite{moore2021ultra, ng15_astro, antoniadis2024second_4, lamb2023rapid, quelquejay2023practical, ellis2024gravitational}.
Such analyses introduce approximations, such as binary orbit circularity, and GW frequencies of sources observed at harmonics of $1/T$, where $T$ is the observation time of the pulsars, which are not accurate.
The same assumptions are also intrinsic to the weighted-Poisson GWB spectral model introduced by \citet{sato2025distribution} (see also \citep{lamb_taylor_24}).
Additionally, a finite observation time introduces frequency covariances that are currently not included in astrophysical inference \citep{crisostomi2025beyond}.

\citet{Xue:2024qtx} presents the distribution of the squared magnitude of the pulsar redshift Fourier coefficient in terms of a semi-analytic model (SAM) of circular SMBHBs.
This work shows how a GWB spectrum distribution can be numerically computed by Fourier transforming the model's characteristic function.
However, they approximate the distribution of observed GW frequencies as a step-function, which does not reflect the predicted power-law distribution for a GW-hardened binary \citep{phinney2001practical, jaffe2003gravitational}.
More recently, \citet{falxa2025modeling} introduces an agnostic Gaussian mixture model to represent the distribution of Fourier coefficients from a PTA time-series to efficiently model non-Gaussianities.
However, it is unclear how this model can be used for downstream astrophysical parameter estimation.

Additionally, these publications do not model the correlated response of two pulsars to a discrete GW source and its associated uncertainty over realizations of sources and pulsars.
For an isotropic, unpolarized, and Gaussian GWB, the mean correlated response between two pulsars is described by the Hellings \& Downs (HD) curve \citep{hellingsdowns}.
However, the correlated response between two pulsars to a GW point source will differ from the mean HD curve in individual realizations of the astrophysical GWB \citep{allen2023variance} (see also \citep{allen_romano2023, allen_valtolina_24, allen:2024uqs}).
Additionally, the finite observation time of an experiment will result in confusion between discrete sources, which will vary over realizations of the GWB.
This is an example of \textit{cosmic variance} \citep{allen2023variance}.

Previous analysis on the astrophysical GWB focuses on the timing-residual power spectral density $S_{\delta t}(f)$ (or quantities related to it, such as the squared characteristic strain $h_c^2(f)$).
However, these quantities implicitly require a Gaussian prior on the Fourier coefficient of the GWB component of the timing residual, with non-Gaussianities introduced as ``corrections'' to this model (see e.g. \citep{Xue:2024qtx}).
In this paper, we investigate the accuracy of the Gaussian prior by extending the simple model of \citet{allen2023variance} to Fourier coefficients of a time-series measured by a pulsar in response to a GWB from a finite population of SMBHBs.
We show that non-Gaussianity in the Fourier coefficient is only expressed in the fourth moment, from which we develop a kurtosis metric to measure said non-Gaussianity and show how Poisson variance, HD variance, and finite observation times affect the moments found in \citet{lamb_taylor_24}.
This work differs from other publications that analyzes, tests, and simulates Gaussianity \citep{bernardo2025toward, bernardo2026accurate, bernardo2026simulating} as our astrophysical model is, by definition, non-Gaussian.
We also analyze the distribution of the argument of the complex Fourier coefficient and show how it can be used to investigate non-Gaussianity.
Finally, we conclude by discussing how our work can be utilized for faster and more accurate simulation and analysis of non-Gaussian GWBs.

Throughout this paper, we will present our model in the context of PTAs and their sources.
However, our framework is applicable across GW detectors, with differences arising from different antenna responses and sources, resulting in different prefactors to our analytical results.

For the busy reader, here is an overview of our results:
\begin{itemize}
\item We derive the Fourier coefficients of the timing residuals from a pulsar $p$ responding to a GWB from a finite population of circular, inclined SMBHBs that are hardened by gravitational wave emission only.
\item We derive the first four moments of the Fourier coefficient for the discrete-source GWB, and discuss different origins of those non-Gaussianities.
\item We investigate the argument of the (complex) Fourier coefficients. The tangent of the argument for a Gaussian, isotropic, and unpolarized GWB has a standard Cauchy distribution, and so looking at the tangent of the argument of the Fourier coefficients provides a complementary probe of non-Gaussianity.
\end{itemize}

The paper is laid out as follows.
In \autoref{sec:astro-gwb}, we introduce the GWB timing-residual model from which we derive Fourier coefficients.
We then find the moments of the coefficients to open \autoref{sec:moments} before looking at the limiting cases of one GW source (\autoref{ssec:single-source-variance}) and many GW sources (\autoref{ssec:many-sources-variance}), and close this section by considering the distribution of the argument of the Fourier coefficients in \autoref{ssec:argument}.
We compare our analytical results to numerical simulations in \autoref{sec:comparing} for both a toy model (\autoref{ssec:toy-model}) and an astrophysical model (\autoref{ssec:astro-model}).
We conclude in \autoref{sec:conclusion}.

Further details on ensemble averaging, complex random variables, window functions, and semi-analytical models can be found in the Appendices.
Throughout, we will use natural units, $c=G=1$.

\section{Observing the Astrophysical Gravitational Wave Background}\label{sec:astro-gwb}

Assuming that all GW sources are far enough from the Earth-pulsar system, we will describe their metric perturbations as plane waves. These plane waves are uniquely characterized by two functions of a single variable, one for each GW polarization, which we call $h^+$ and $h^\times$:
\begin{widetext}\begin{subequations}\begin{align}
    h_j^+(t) &= A_j\left[\frac{1}{2}\left(1+\cos^2\iota_j\right)\cos2\psi_j\cos\left(2\pi f_jt+\phi_j\right) - \cos\iota_j\sin 2\psi_j \sin\left(2\pi f_jt+\phi_j\right)\right]\,, \\
    h_j^\times(t) &= A_j\left[\frac{1}{2}\left(1+\cos^2\iota_j\right)\sin2\psi_j\cos\left(2\pi f_jt+\phi_j\right) + \cos\iota_j\cos 2\psi_j \sin\left(2\pi f_jt+\phi_j\right)\right]
\end{align}\label{eq:finite-strain}\end{subequations}\end{widetext}
where the amplitude of the metric perturbation, $A(f_j, \vec{\theta}_j)\equiv A_j$, depends on the astrophysical parameters and frequency of its source and we have assumed that the SMBHB orbit is circular (from which follows that the relative phase between the two polarizations is $\pi/2$). The inclination $\iota_j$ of the binary is defined as the angle between the line of sight and the orbital angular momentum, with $\cos\iota_j\in[-1, 1]$. Inclined, circular orbits appear elliptical to the observer, and its orientation is described by the polarization angle $\psi_j\in[0, \pi/2]$. The rest-frame frequency $f_j$ is related to the observed GW frequency $f_{j,\mathrm{obs}}$ by $f_j=(1+z_j)f_{j,\mathrm{obs}}$, where $z_j$ is the redshift between the observer and the source. The metric perturbation produced by the entire population of sources at time $t$ and location $\vec{x}$ is then given by:
\label{eqn:hab-1}\begin{align}
    h_{ab}(t, \vec{x}) &= \sum_{j=1}^{N} \sum_{A\in\{+,\times\}} h_j^A(t-\hat\Omega_j\cdot\vec x)\epsilon^A_{ab}(\hat{\Omega}_j).
\end{align}

The timing residual for the $p$-th pulsar induced by this metric perturbation is given by:
\begin{align}\label{eq:gaussian_gwb_amp}
    \delta t_p(t) &= \frac{\hat{x}_p^a \hat{x}_p^b}{2}\int^t_{t-L_p} dt' h_{ab}[t', (t-t')\hat{x}_p] 
\end{align}
where $\hat{x}_p$ is the unit vector pointing from Earth to the $p$-th pulsar, and $L_p$ is the distance from the Earth to the pulsar.
We define a complex antenna response function: $F_{p,j} \equiv F_p^+(f_j, \hat\Omega_j) - iF_p^\times(f_j, \hat\Omega_j)$, where the antenna response function for pulsar $p$ is given by:
\begin{subequations}\begin{align}
    F_p^A(\hat\Omega) &= \frac{1}{2}\frac{\hat{x}^a_p \hat{x}^b_p}{1+\hat{\Omega}\cdot\hat{x}_p}\epsilon^A_{ab}(\hat{\Omega}).
\end{align}\end{subequations}

We also define the Earth-pulsar response factor,
\begin{equation}
    \chi_{p,j} = 1-e^{-2\pi ifL_p(1+\hat{\Omega}\cdot \hat{x}_p)},
\end{equation}
which encodes the relative (and delayed) responses of the Earth and pulsar in the presence of a gravitational wave.
It is common in PTA analyses to decompose the timing residual on a discrete, \textit{two-sided} Fourier basis with sampling frequencies $f_i=i/T$, where $T$ is the total observation time:
\begin{equation}
    \delta t_p(t) = \sum_i \tilde{a}^p_i e^{2\pi if_i t}\,,
\end{equation}
where the summation is usually truncated at some frequency between the largest GWB-dominated frequency and the inverse of the observational cadence. Then, for our population model, the Fourier coefficient for the $i$-th frequency mode is
\begin{widetext}\begin{align} \label{eq: finite pop fourier coeff}
    \tilde{a}_i^p = \frac{1}{8}\sum_j A_j \left[\frac{w^-_j}{2\pi if_j} \left(\left(1+\cos\iota_j\right)^2e^{2i\psi_j}F_j + \left(1-\cos\iota_j\right)^2e^{-2i\psi_j}F^*_j\right) \chi_{p,j}\,e^{i\phi_j}+ \mathrm{c.c}\right]
\end{align}\end{widetext}
where we have introduced the window function, $w_j^\pm\equiv w(f_i\pm f_j)$. For an observation carried out for a finite time $T$, the window function takes the form $w_j^\pm={\rm sinc}[\pi(f_i\pm f_j)T]$. However, in a realistic PTA observation, the window function will be modified by the marginalization over the timing model parameters. In this work, we will derive analytical results without specifying the form of the window function, but only assuming that they are real functions. However, to simplify our equations, we will consider $w_j^+$ as the ``complex conjugate'' of $w_j^-$. Therefore, our final results can be applied to different forms of the window function.

In PTA analyses, the Fourier coefficients of the timing residuals are typically assumed to be a Gaussian-distributed complex random variable, reflecting the idea that the GWB itself is Gaussian. In the next sections, we will find expressions for the higher moments of $\tilde{a}^p_i$ to test this Gaussian assumption.

\section{Moments of the Finite-Population Gravitational Wave Background}\label{sec:moments}
In this section, we derive expressions for the moments of the timing-residual Fourier coefficients, $\tilde{a}_i^p$, generated by the finite population model discussed in the previous section. To aid the pedagogy of our derivation, we begin by deriving the moments on $\tilde{a}_i^p$ for the case of a ``face-on,'' circular binary system with inclination angle $\iota=0$ (or $\iota=\pi$) and orientation angle $\psi_j\equiv0$. This system is called `unpolarized' \citep{conneely2019amplitude}.  Hence, \autoref{eq: finite pop fourier coeff} reduces to,
\begin{align} \label{eq: circular fourier coeff}
    \tilde{a}_{i,\mathrm{unpol}}^p = \sum_j \frac{\mathcal{A}_j}{2}\left(\frac{w_{j}^-}{2\pi i f_j}e^{i\phi_j}R_{p,j} + \mathrm{c.c.} \right)\,,
\end{align}
where we have defined $R_{p,j}\equiv\chi_{p,j}F_{p,j}$. Following \citet{allen_valtolina_24}, we also redefine the amplitude from $A_j$ to $\mathcal{A}_j$. We will show that, to achieve the same second moment between the inclination- and polarization-averaged `unpolarized' model, and the general `polarized' model that allows inclination and polarization angle to vary, then $\mathcal{A}_j=\sqrt{2/5}A_j$.

We start by computing $\langle\tilde{a}_{i,\mathrm{unpol}}^p\rangle$, where $\langle\cdot\rangle$ denote the ensemble average. In our model, the ensemble average consists of averaging over phases, $\phi_j$, source positions, $\hat\Omega_j$, and the number of sources. We start by computing the average over phases, which is defined as:
\begin{equation}
    \langle Q\rangle_\phi\equiv\int_0^{2\pi} \frac{d\phi_1}{2\pi}\ldots\int_0^{2\pi} \frac{d\phi_N}{2\pi}\;Q(\phi_1,\ldots,\phi_N)\,.
\end{equation}
Using the standard result, $\langle e^{i\phi}\rangle_\phi=0$, we immediately find that
\begin{equation}
    \left<\tilde{a}_{i,\mathrm{unpol}}^p\right>_\phi = 0\,,
\end{equation}
In fact, all odd moments of $\tilde{a}_{i,\mathrm{unpol}}^p$ are zero mean because they have an odd number of $e^{i\phi_j}$ components. Since $ \left<\tilde{a}_{i,\mathrm{unpol}}^p\right>_\phi$ does not depend on the source positions or properties, the phase average coincides with the ensemble average: $ \left<\tilde{a}_{i,\mathrm{unpol}}^p\right>_\phi= \left<\tilde{a}_{i,\mathrm{unpol}}^p\right>\equiv0$. By extension, for the general `polarized' model, $ \left<\tilde{a}_{i}^p\right>_\phi= \left<\tilde{a}_{i}^p\right>\equiv0$.

We now move our attention to the second moment of timing residuals' Fourier components. We start by computing the cross-pulsar product for pulsars $p$ and $q$,
\begin{align}\label{eq:finite_pop_product}
    \tilde{a}_{i,\mathrm{unpol}}^{*p}\tilde{a}_{i,\mathrm{unpol}}^{q} &= \sum_j \frac{\mathcal{A}_j^2}{16\pi^2 f_j^2}\left[w_{j}^{--} R_{pj}^*R_{qj} + w_{j}^{++} R_{pj}R^*_{qj} \right. \nonumber \\
    &\qquad\qquad\qquad- \left. w_{j}^{-+} \left(R_{pj}R_{qj} e^{2i\phi_j} + \text{c.c.}\right)\right] \nonumber \\
    &+ \sum_{j\neq k} \frac{\mathcal{A}_j \mathcal{A}_k}{16\pi^2 f_j f_k} \left(w_{j}^- R^*_{pj} e^{-i\phi_j} - w_{j}^+ R_{pj} e^{i\phi_j}\right) \nonumber \\
    &\qquad\qquad\times \left(w_{k}^- R_{qk} e^{i\phi_k} - w_{k}^+ R^*_{qk} e^{-i\phi_k}\right),
\end{align}
where $j,k$ are indices over the sources, and we have defined $w^{\pm\pm} _j = w(f_j\pm f_i)w(f_j\pm f_i)$. Taking the auto-pulsar product as the case where $p=q$, we find the phase average as
\begin{equation}
    \left\langle\left|\tilde{a}_{i,\mathrm{unpol}}^p\right|^2\right\rangle_\phi = \sum_j \frac{\mathcal{A}_j^2}{16\pi^2 f_j^2}\left|R_{pj}\right|^2\left(w_{j}^{--} + w_{j}^{++}\right)\,.
\end{equation}
The above phase average still depends on the source positions via the $|R_{pj}|^2$ term. Therefore, to find the ensemble average, we perform an average over all possible source directions. For PTAs, the sky average of the product of two response functions is related to the well-known Hellings-\&-Downs curve, $\Gamma(\gamma)$, \citep{hellingsdowns}:
\begin{align}
    \Gamma_{pq}\equiv\Gamma(\gamma_{pq}) = \frac{3}{2}\left<R^*_{pj}R_{qj}\right>_{\hat\Omega}\,,
\end{align}
where $\gamma_{pq}$ is the angular separation between pulsar $p$ and pulsar $q$, and the normalization factor of $3/2$ is chosen such that the auto-correlation $\Gamma_{pp}=1$. From this, it follows that:
\begin{equation}\label{eqn:a-variance}
    \left<\left|\tilde{a}_{i,\mathrm{unpol}}^{p}\right|^2\right>_{\phi, \hat\Omega} = \sum_j \frac{\mathcal{A}_j^2}{24\pi^2 f_j^2} \left(w_{j}^{--} + w_{j}^{++} \right)\,.
\end{equation}
While we have only reported the second moment for the $p=q$ case, the cross-pulsar second moment is very similar, and it can be obtained by multiplying the right-hand side of \autoref{eqn:a-variance} by $\Gamma_{pq}$.

Following an analogous procedure (i.e., averaging over phases and sky positions -- for a detailed derivation, see \aref{app:ensemble}), we find that the four-point function is given by
\begin{widetext}\begin{equation}\label{eq:4th-poisson-fixed}
    \left<|\tilde{a}_{i,\mathrm{unpol}}^p|^4\right>_{\phi,\hat\Omega} = 2\left<|\tilde{a}_{i,\mathrm{unpol}}^p|^2\right>_{\phi,\hat\Omega}^2
    +0.7\sum_j \frac{\mathcal{A}_j^4}{24^2\pi^4 f_j^4} \left[(w^{--}_j + w^{++}_j)^2 + 2w^{--}_jw^{++}_j\right]
    + 4\left(\sum_j \frac{\mathcal{A}_j^2}{24\pi^2 f_j^2}w^{-+}_j\right)^2 \,.
\end{equation}
\end{widetext}
The factor of 0.7 comes from the fourth moment of the antenna response, $\left<|R_{pj}|^4\right>$. To see how this arises, we first set $\gamma_{pq}=0$ in \autoref{eq:cross-pol} to obtain $2.7$. Then, the difference of two is found in the first term of the right-hand side of \autoref{eq:4th-poisson-fixed} as the diagonal of the squared term.

Next, we will average over fluctuations in the number of sources. To do this, we divide the parameter space of the SMBHB model into equal-size bins, and replace the sum over individual sources in \autoref{eqn:a-variance} with a sum over these bins:
\begin{equation}\label{eq:sum_pver_bins}
    \sum_{j=1}^N \frac{\mathcal{A}_j^2}{f_j^2} \quad\Rightarrow\quad\sum_J N_J \frac{\mathcal{A}_J^2}{f_J^2}\,,
\end{equation}
where $N_J$ is the number of sources in the $J$-th bin with $\sum_J N_J = N$, and $A_J$, $f_J$ are the amplitude and frequency of the sources in that bin. The number of sources in each bin is a Poisson random variable, with a mean value $\bar N_J$: $N_J \sim \mathcal{P}\left(\bar{N}\right)$. This mean value of sources can be derived, for a given set of astrophysical parameters, using semi-analytical models (SAM) or numerical simulations. SAMs and the process of ensemble averaging over $N_J$ is detailed in \aref{app:sam}. After averaging over $N_J$, we get:

\begin{align}\label{eqn:a-variance-N}
    \left<\left|\tilde{a}_{i,\mathrm{unpol}}^{p}\right|^2\right> &\equiv \left<\left|\tilde{a}_{i,\mathrm{unpol}}^{p}\right|^2\right>_{\phi, \hat\Omega, N}\,\nonumber\\
    &= \sum_J \bar N_J\frac{\mathcal{A}_J^2}{24\pi^2 f_J^2} \left(w^{--}_J + w^{++}_J \right)\,.
\end{align}
\autoref{eqn:a-variance-N} is equivalent to discrete SAMs of GWBs used in astrophysical inference of PTA data \citep{Sesana:2008mz, ng15_astro}, with the addition of window functions from finite observation times. Finally, after averaging over $N_J$, we find that the four-point function is given by:
\begin{widetext}\begin{equation}\label{eq:4th-poisson}
    \left<|\tilde{a}_{i,\mathrm{unpol}}^p|^4\right>_{\phi,\hat\Omega,N} = 2\left<|\tilde{a}_{i,\mathrm{unpol}}^p|^2\right>^2
    +2.7\sum_J \bar N_J \frac{\mathcal{A}_J^4}{24^2\pi^4 f_J^4} \left[(w^{--}_J + w^{++}_J)^2 + 2w^{--}_Jw^{++}_J\right]
    + 4\left(\sum_J\bar N_J\frac{\mathcal{A}_J^2}{24\pi^2 f_J^2}w^{-+}_J\right)^2 \,,
\end{equation}
\end{widetext}
where an additional factor of $2$ in the second term comes from assuming Poisson statistics when averaging over the first term of the right-hand side (\autoref{eq:poisson_definition}), resulting in the factor of $2.7$.

We observe that, in the large number limit $\bar{N}_J\gg1$, the first and third term on the right-hand side of \autoref{eq:4th-poisson} is significantly greater than the second term, since,
\begin{equation}\label{eq:bigger_gaussian}
    \sum_J \bar N_J \mathcal{A}_J^4 \ll \left(\sum_J \bar N_J \mathcal{A}_J^2\right)^2\,\mathrm{for}\, \bar{N}_J\gg1.
\end{equation}
In fact, in the limit $\sum_J \bar{N}_J\to\infty$, \autoref{eq:4th-poisson} is equivalent to the fourth moment of a complex Gaussian random variable (see \autoref{eq:isserlis}). We will revisit this point later on, but we use this relation to interpret \autoref{eq:4th-poisson} as the fourth moment of a complex Gaussian random variable with an additional correction term that results from Poissonian uncertainty in the finite number of sources per realization. We use this information to bootstrap our way to computing the moments of the `polarized' model $\tilde{a}_i^p$ of \autoref{eq: finite pop fourier coeff}, where we introduce inclination and polarization angle. We start by computing the second moment of $\tilde{a}_i^p$ (\autoref{eqn:a-variance-N}):
\begin{align}\label{eqn:a-variance-N-pol}
    \left<\left|\tilde{a}_{i}^{p}\right|^2\right> = \frac{2}{5} \sum_J \bar N_J\frac{A_J^2}{24\pi^2 f_J^2} \left(w^{--}_J + w^{++}_J \right)\,.
\end{align}
The factor of $2/5$ arises from averaging over inclination angle. As discussed in \citet{allen_valtolina_24}, \autoref{eqn:a-variance-N-pol} is equivalent to \autoref{eqn:a-variance-N} if $\mathcal{A}_j^2= 2A_j^2/5$.

For the fourth moment of $\tilde{a}_{i}^{p}$, its structure is identical to \autoref{eq:4th-poisson} -- the sum of the square of the second moment, a Poisson term, and a covariance term -- albeit with different coefficients. For the terms that sum over $A_j^2$, we simply carry over the factor of $2/5$ that results from ensemble averaging over inclination angle. The coefficient for Poissonian term is more involved and derived in \aref{app:ensemble}. Hence,

\begin{widetext}\begin{align}\label{eq:4th-poisson-pol}
    \left<|\tilde{a}^p_{i}|^4\right> = 2\left<|\tilde{a}^p_{i}|^2\right>^2 + \frac{108}{175}\sum_J \bar N_J \frac{A_J^4}{24^2\pi^4 f_J^4} \left[(w^{--}_J + w^{++}_J)^2 + 2w^{--}_Jw^{++}_J\right]
    + 4\left(\frac{2}{5}\sum_J\bar N_J\frac{A_J^2}{24\pi^2 f_J^2}w^{-+}_J\right)^2 \,.
\end{align}\end{widetext}

\autoref{eq:4th-poisson-pol} is one of the main results of this work, so let us discuss it in more detail. First, we immediately notice that $\tilde{a}_{i}^p$ is non-Gaussian. This non-Gaussianity, which arises from a combination of finite-source effects and Poisson fluctuations in the total number of GW sources, is quantified by computing the excess kurtosis, $\bar\kappa\equiv \langle|\tilde a_i^p|^4\rangle/\langle|\tilde a_i^p|^2\rangle -2$. For a circularly complex Gaussian random variable, $\bar\kappa=0$ (see \autoref{app:complex-rvs} for more details), while from \autoref{eq:4th-poisson-pol} we find that the excess kurtosis, $\bar\kappa$, is
\begin{widetext}
\begin{align}\label{eq:ex_kurt}
    \bar\kappa
    \simeq
    \frac{
        \frac{27}{7}\sum_{J} \bar{N}_J A_J^{4} \left[\left(w_J^{--} + w_J^{++}\right)^2 + 2{w_J^{-+}}^2\right] /f_J^{4}
        + 4 \left( \sum_{J} \bar{N}_J A_J^{2} w_J^{-+} / f_J^{2} \right)^{2}
    }{
        \left[\sum_{J} \bar{N}_J A_J^{2} \left(w_J^{--}+w_J^{++}\right) / f_J^{2} \right]^{2}
    }.
\end{align}
\end{widetext}
Often, in astrophysical inference analyses, the sources are assumed to be emitting only at frequencies that are integer multiples of the observing time. In this case, $w^{--}=1$ while $w^{++,+-}=0$, and the excess kurtosis reduces to:
\begin{align}\label{eq:ex_kurt_simple}
    \bar\kappa
    \simeq
    \frac{27}{7} \frac{
        \sum_{J} \bar{N}_J A_J^{4}  /f_J^{4}
    }{
        \left( \sum_{J} \bar{N}_J A_J^{2}/ f_J^{2} \right)^{2}
    }\,.
\end{align}
In the case where we generate realizations of a fixed number of sources $N$, and do not average over some distribution of $N$, we subtract $2$ from the coefficient that results from assuming Poisson statistics (\autoref{eq:poisson_definition}):
\begin{align}\label{eq:ex_kurt_N}
    \bar\kappa_N
    \simeq
    \frac{13}{7}
    \frac{
        \sum_{j}^N A_j^{4} /f_j^{4}
    }{
        \left( \sum_{j}^N A_j^{2} / f_j^{2} \right)^{2}
    }\,,
\end{align}
\subsection{Single-source limit}\label{ssec:single-source-variance}
It is also instructive to study the form of our results in two opposite limits: realizations of an individual source; and realizations of populations where the mean number of sources $\bar{N}\gg1$. If each realization features only one source with amplitude $A$ and frequency $f$, the two- and four-point functions take the form
\begin{subequations}
    \begin{align}
    \left<|\tilde{a}_i^p|^2\right>_{\phi,\hat\Omega}&=\frac{2}{5}\frac{A^2}{24\pi^2f^2}w^{--}\,,\\
    \left<|\tilde{a}_i^p|^4\right>_{\phi,\hat\Omega}&\simeq2\left<|\tilde{a}_i^p|^2\right>^2_{\phi,\hat\Omega} + \frac{13}{7}\left<|\tilde{a}_i^p|^2\right>^2_{\phi,\hat\Omega}\label{eq:4th-single}\,,
\end{align}
\end{subequations}
where we have ignored all terms containing a $w^{++}$ or $w^{+-}$, since we are working under the assumption that they are much smaller compared to $w^{--}$. So it is easy to see from \autoref{eq:ex_kurt_N} that, in this limit, the excess kurtosis is $\bar\kappa_{N=1}\simeq1.86$; a result that we will explicitly check with the numerical simulations discussed in \autoref{sec:comparing}.

\subsection{Many-sources limit}\label{ssec:many-sources-variance}

In the opposite limit, when many, but an unknown number, of sources contribute to the GWB such that $\bar{N}_J\gg1$ for the majority of the parameter-space bin, the four-point function of \autoref{eq:4th-poisson} reads
\begin{equation}\label{eq:auto_many}
    \left<|\tilde{a}_i^p|^4\right> \simeq 2\left<|\tilde{a}_i^p|^2\right> + 
    4\left(\frac{2}{5}\sum_J\bar N_J\frac{A_J^2}{24\pi^2 f_J^2}w^{-+}\right)^2\,,
\end{equation}
where we have ignored the second term on the right-hand side of \autoref{eq:4th-poisson-pol} given \autoref{eq:bigger_gaussian} for $\bar{N}_J\gg1$. Even in the many-sources limit, there is a residual non-Gaussian term. This residual non-Gaussianity derives from the covariance between the real and imaginary parts of $\tilde a_i^p$ introduced by the windowing of our data -- even for a perfect GWB, fully described by the two-point function of the $\tilde{h}_A(f, \hat{\Omega})$ coefficients in \autoref{eq:gaussian_gwb_corr}. The second and fourth moments of the Fourier coefficients are
\begin{subequations}\begin{align}
   \left<|\tilde{a}_i^{p}|^2\right> &= \int_{-\infty}^\infty \frac{S_h(f)}{24 \pi^2 {f}^2} \left(w^{--}+w^{++}\right) df \label{eq:gauss_2nd_moment}\,,\\
    \left<|\tilde{a}_i^p|^4\right> &= 2\left<|a_i^p|^2\right>^2 + \left( \int_{-\infty}^\infty \frac{S_h(f)}{24\pi^2f^2}w^{-+} df \right)^2\,.\label{eq:gauss_4th_moment}
\end{align}\end{subequations}
Note that the second term in \autoref{eq:gauss_4th_moment} is comparable to the second term in \autoref{eq:auto_many} -- it is a covariance term between the real and imaginary components of a Gaussian random variable. The difference of a factor of 4 is a result of our integration limits and definition of the (two-sided) GWB strain PSD $S_h(f)$, while the factor of $2/5$ is absorbed into $S_h(f)$ as it is an inclination-averaged quantity. The integration limits of \autoref{eq:auto_many} are between $0$ and $\infty$ because discrete sources have positive frequencies. If we change the integration limits in the second term in \autoref{eq:gauss_4th_moment} such that the PSD is one-sided, it would introduce a factor of $4$. Hence, the many-source limit of the astrophysical GWB may be approximated as a Gaussian GWB. 

\subsection{Argument of the finite-population Fourier coefficient}\label{ssec:argument}

Since $\tilde{a}_i^p$ is a complex random variable, we are able to analyze its argument using \autoref{eq:complex_arg}. In this paper, we will only analytically investigate the argument of the `unpolarized' model, $\tilde{a}_{i,\mathrm{unpol}}^p$. Taking the real and imaginary components of \autoref{eq: circular fourier coeff}, using trigonometric identities, assume that $f_jL_p\gg1$ such that $R_{pj}\approx F_{pj}$, and substituting $\phi_j'=\phi_j-\pi/2$, we find that
\begin{subequations}
\begin{align}
    &\mathrm{Re}\left[\tilde{a}_{i,\mathrm{unpol}}^p\right] = \\
    &\sum_j \frac{\mathcal{A}_j}{4\pi f_j}\left(w_{j}^- + w_{j}^+\right)\left[F_{pj}^+ \cos\phi_j' + F_{pj}^\times \cos\left(\phi_j'-\frac{\pi}{2}\right)\right] \nonumber \\
    &\mathrm{Im}\left[\tilde{a}_{i,\mathrm{unpol}}^p\right] = \\ 
    &\sum_j \frac{\mathcal{A}_j}{4\pi f_j}\left(w_{j}^- - w_{j}^+\right)\left[F_{pj}^+ \sin\phi_j' + F_{pj}^\times \sin\left(\phi_j'-\frac{\pi}{2}\right)\right] \nonumber
\end{align}
\end{subequations}
Hence, the argument $\varphi_i^p$ \textit{of a single realization} of the Fourier coefficient measured by pulsar $p$ at observation frequency $f_i$ is,
\begin{align}\label{eq:finite-cauchy}
    &\tan\varphi^p_i = \\
    &\qquad\frac{\sum_j \frac{\mathcal{A}_j}{ f_j}\left(w_{j}^- - w_{j}^+\right)\left[F_{pj}^+ \sin\phi_j' + F_{pj}^\times \sin\left(\phi_j'-\frac{\pi}{2}\right)\right]}{\sum_j \frac{\mathcal{A}_j}{ f_j}\left(w_{j}^- + w_{j}^+\right)\left[F_{pj}^+ \cos\phi_j' + F_{pj}^\times \cos\left(\phi_j'-\frac{\pi}{2}\right)\right]}. \nonumber
\end{align}
Without loss of generality, we locate the pulsar $p$ parallel to the z-plane such that $F^\times(f,\hat\Omega)=0$. In the case where $N=1$, then,
\begin{align}\label{eq:cauchy-finite}
    &\tan\varphi^p_i = \frac{\left(w_{j}^- - w_{j}^+\right)}{\left(w_{j}^- + w_{j}^+\right)} \tan\phi_j',
\end{align}
Therefore, if $\tan\phi'_j$ follows a Cauchy distribution with location and scale parameters $x_{0,j}$ and $\gamma_j$, then \autoref{eq:cauchy-finite} is a Cauchy distribution with location $x_{0,j}'=\beta_j x_{0,j}$ and scale $\gamma_j'=\beta_j\gamma_j$, where
\begin{equation}
    \beta_j=\frac{w^-_{j} - w^+_{j}}{w^-_{j} + w^+_{j}}.
\end{equation}
Future studies can introduce eccentric single sources, in which case the equal-amplitude assumption of \autoref{eq:finite-strain} is invalid and the distribution of $\tan\varphi_i^p$ for a single source may no longer be a Cauchy distribution. Regardless, in the large source limit, the distribution of $\tan\varphi_i^p$ will tend toward Cauchy. More details on Cauchy random variables may be found in \aref{app:cauchy}.

\section{Comparing to numerical results}\label{sec:comparing}

In this section, we use numerical simulations to test the analytical results presented in \autoref{sec:moments}. Specifically, we will use these simulations to estimate the moments of timing residuals Fourier coefficients of a single pulsar, which we will term the \emph{auto-pulsar moments}, as well as the cross-moments of Fourier coefficients of two pulsars, which we will term \emph{cross-pulsar moments}.

For cross-pulsars, the second moment is the ensemble average of the product of two Fourier components, $\langle\tilde{a}^{*p}_i\tilde{a}^q_i\rangle$, while the fourth moment is the absolute square of this product, $\langle|\tilde{a}^{*p}_i\tilde{a}^q_i|^2\rangle$. An important result in the large-source-number limit is that the cross-pulsar fourth moment tends towards twice the second moment of \emph{either} pulsar individually: $\langle|\tilde{a}^{*p}_i\tilde{a}^q_i|^2\rangle\approx 2\langle|\tilde{a}^p_i|^2\rangle^2$.
    
\subsection{Toy Model with Fixed Number of Sources}\label{ssec:toy-model}

We start by considering the simple model of SMBHB (first described in~\citep{allen2023variance}) where we have a fixed number of sources, $N$, and the amplitude $\mathcal{A}_j$ of the $j^\mathrm{th}$ source is given by $\mathcal{A}_j = \mathcal{A}j^{-\frac{1}{3}}$ (where $\mathcal{A}$ is an arbitrary amplitude). Moreover, we assume that all sources emit at frequencies which are integer multiples of the inverse of the observing time, such that the window functions are $w^-_j,w^+_j=1,0$. 
We then generate $10^5$ realizations of this toy model, and for each of these realizations, we use \autoref{eq: finite pop fourier coeff} to compute the timing residuals Fourier coefficients. We then use these coefficients to estimate the excess kurtosis.

\begin{figure}
    \centering
    \includegraphics[width=1.0\linewidth]{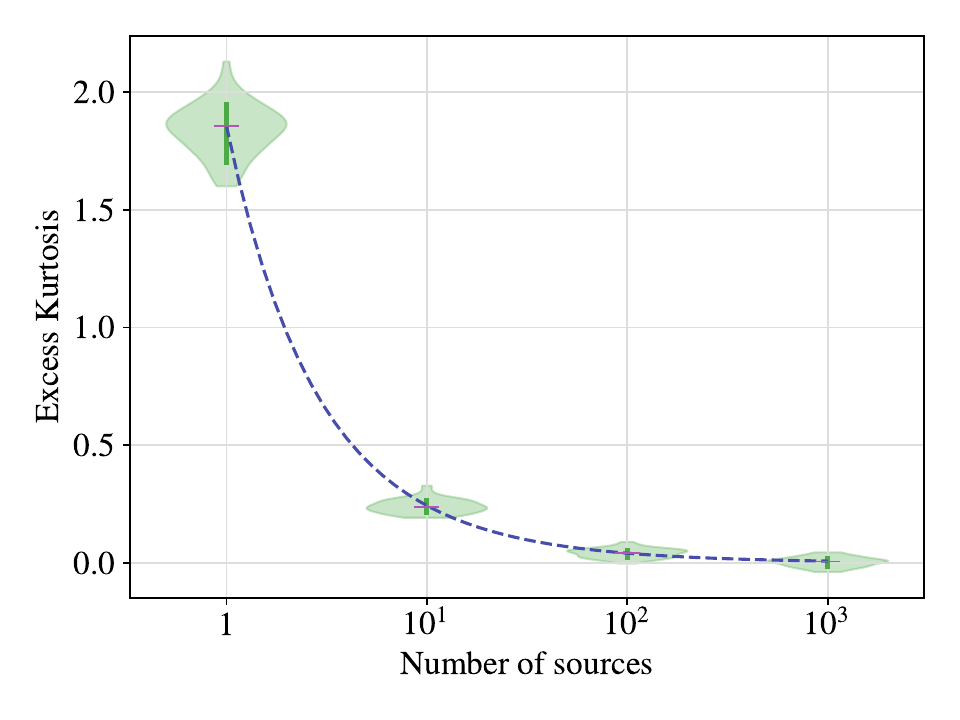}
    \caption{Comparing excess kurtosis as a function of the number of sources numerically (green violin plots) and analytically (dashed blue curve). The vertical green bar indicates the 10\% to 90\% quantile range of the distribution of the excess kurtosis estimator across the $67$ pulsars in the array; the horizontal pink bar indicates the median. The kurtoses are generated from $10^5$ realizations of an SMBHB population with $f_i=f_j=2/T$.}
    \label{fig:ex-kurt-test}
\end{figure}

In \autoref{fig:ex-kurt-test}, we show the excess kurtosis as a function of the number of sources obtained in this way. These results are compared with the analytical predictions obtained in \autoref{eq:ex_kurt_N}, which for this toy model takes the form
\begin{equation}\label{eq:exkurt-toy}
    \bar\kappa_N = \frac{13}{7}\frac{\sum^N_j j^{-4/3}}{\left(\sum^N_j j^{-2/3}\right)^2}\,.
\end{equation}
We can see that numerical and analytical results show good agreement, and for $N=1$, the excess kurtosis is $\sim1.86$, as predicted. We also see that as $N\to\infty$, then $\bar\kappa_N\to0$. This is expected because as $N\to\infty$, the numerator of \autoref{eq:exkurt-toy} $\to\zeta\left(\frac{4}{3}\right)$, where $\zeta$ is the Riemann zeta function. This is finite. However, the denominator diverges, hence the excess kurtosis goes to zero as $N\to\infty$.\footnote{It is tempting to state that the denominator of \autoref{eq:exkurt-toy} is $\zeta\left(\frac{2}{3}\right)^2$ for $N\to\infty$. However, this would not be correct because the Riemann zeta function $\zeta(s)$ is defined for $\mathrm{Re}[s] > 1$, with its \textit{analytic continuation} used to extend the domain.}
    
\begin{figure}
    \centering
    \includegraphics[width=1.0\linewidth]{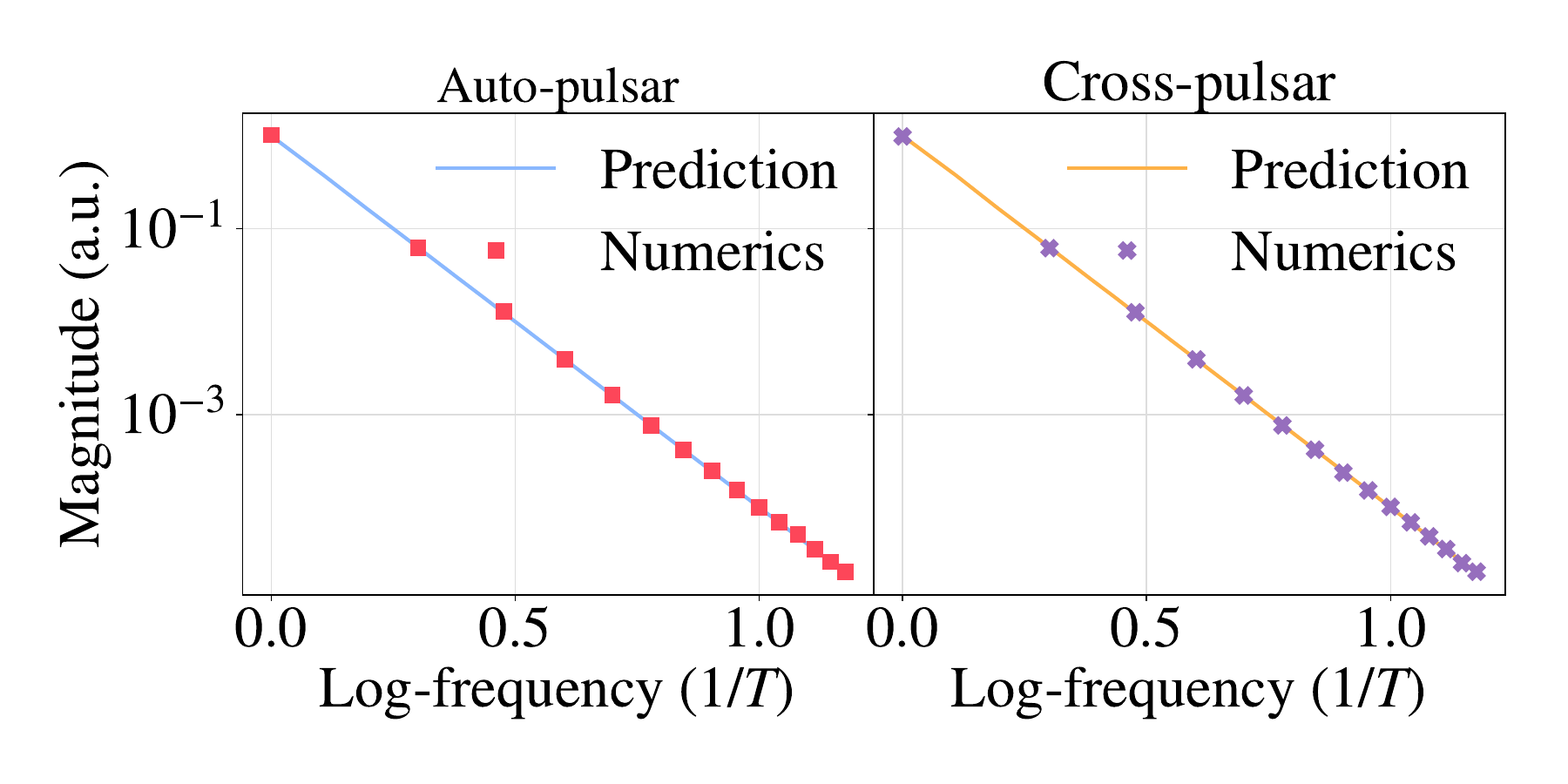}
    \caption{To validate our analytic approach, we compare the predicted fourth moments of the Fourier coefficients (\autoref{eq:4th-single} and its cross-pulsar generalization) with the results of a toy model simulation for a single SMBHB source. The data shown here are generated from $10^7$ realizations for frequencies from $1/T$ to $15/T$. The vertical axes have been rescaled so that the predicted value at $f=1/T$ is $1$ in arbitrary units in both cases.}
    \label{fig:auto-cross_single-source}
\end{figure}

Next, we consider the predictions made for how the second and fourth moments of Fourier coefficients are related. We test the case of a single SMBHB source, in \autoref{fig:auto-cross_single-source}, which validates the frequency scaling of \autoref{eq:4th-single} and its cross-pulsar generalization. As the base amplitude $A$ is arbitrary, we have rescaled the axes so that the predicted value at the lowest frequency is equal to unity.
    
\begin{figure}
    \centering
    \includegraphics[width=1.0\linewidth]{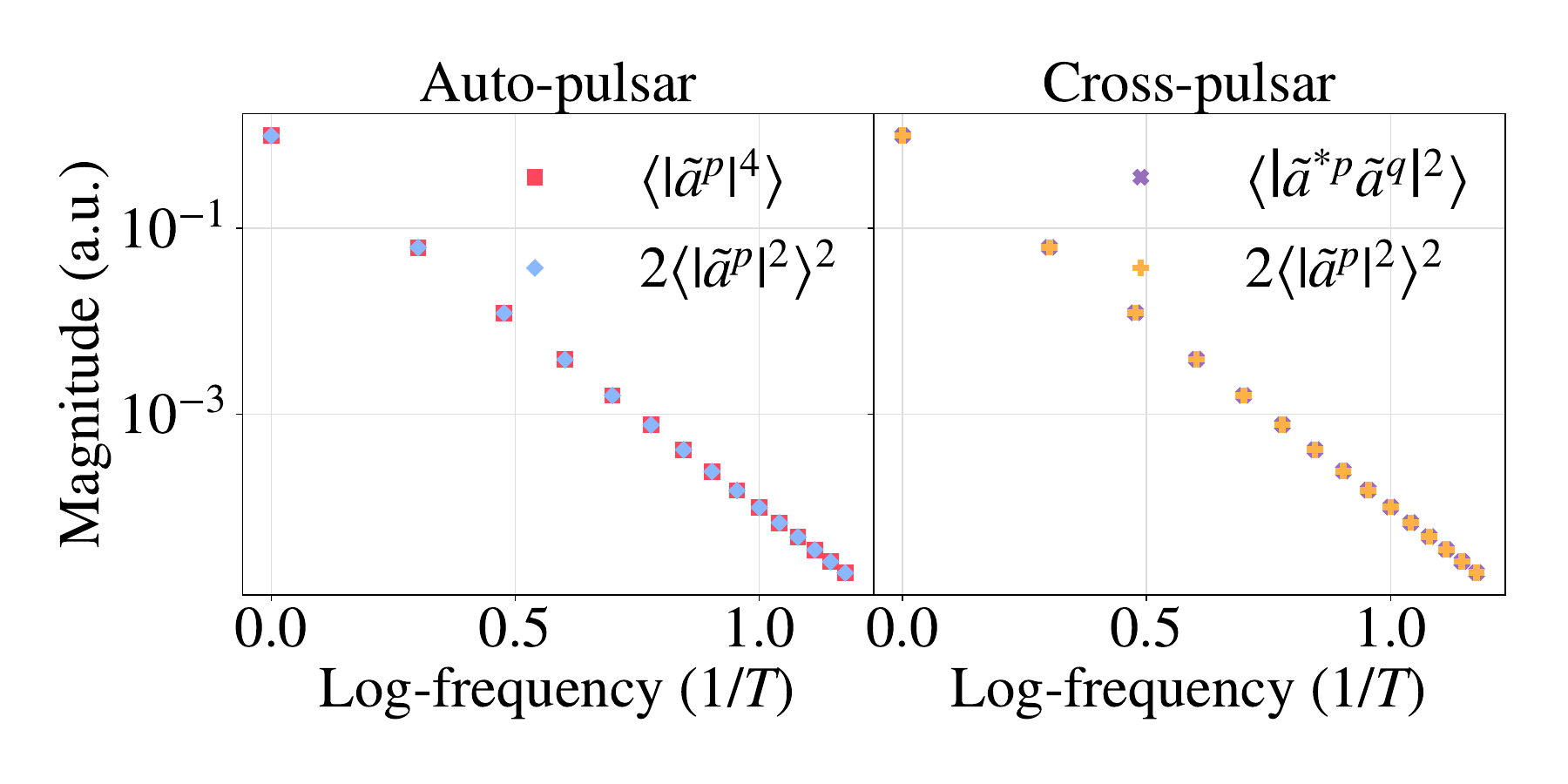}
    \caption{To validate our analytic approach, we compare the predicted fourth moments of the Fourier coefficients (\autoref{eq:auto_many},\autoref{eq:4th-moment-cross}) with the results of a toy model simulation for many SMBHB sources. The data shown here are generated from $10^7$ realizations of an SMBHB population of 1000 sources for frequencies from $1/T$ to $15/T$. The vertical axes have been rescaled so that $\langle|\tilde{a}_i^p|^4\rangle=1$ or $\langle|\tilde{a}_i^{*p}\tilde{a}_i^q|^2\rangle=1$, as appropriate, in arbitrary units at $f=1/T$.}
    \label{fig:auto-cross_many-sources}
\end{figure}

We test the case of many sources (here, $N=1000$) in \autoref{fig:auto-cross_many-sources} to validate \autoref{eq:auto_many} and the cross-pulsar extension, given in \autoref{eq:4th-moment-cross}. In this case, we have no analytic functions to compare to; instead, we compare the variance and the square of the mean directly. As before, the averaging is done over $10^7$ realizations, and the normalization of the fourth moment at $f=1/T$ to unity is done to remove the arbitrary base amplitude.

\subsection{Astrophysical Toy Model}\label{ssec:astro-model}

Let us now consider a more realistic toy model in which the number of sources in each realization is also allowed to vary. We generate data for a total observing time of $T=15\,\text{yr}$ and the 67 pulsars from the NANOGrav 15yr dataset~\cite{agazie2023nanograv_timing}.
Our astrophysical toy model features characteristics that are motivated by true astrophysical models e.g., \citet{middleton2015astrophysical}. We define a simple strain amplitude $A_J$ at each grid cell $J$, with $A_J=N_JM_Jf_J^{2/3}$. Our 2D grid of amplitudes is defined along a coarse grid of frequencies $f_J$ and a random variable $M_J$. The frequency $f_J$ of the sources in cell $J$ are drawn from a power-law distribution with spectral index $-11/3$, as expected for a population of circular SMBHBs inspiralling due to gravitational-wave emission only \citep{peters1964gravitational}. The random variable $M_J$ is drawn from a Rayleigh distribution. The power of $2/3$ is motivated by the strain amplitude of a gravitational wave $h\propto f^{2/3}$. We define the total mean number of sources across our grid as $\sum_J\bar{N}_J=50,000$. Therefore, the mean number of sources $\bar{N}_J$ in cell $J$ is weighted by the probability densities of $f_J$ and $M_J$ in that particular cell.

\begin{figure}
    \centering
    \includegraphics[width=1.0\linewidth]{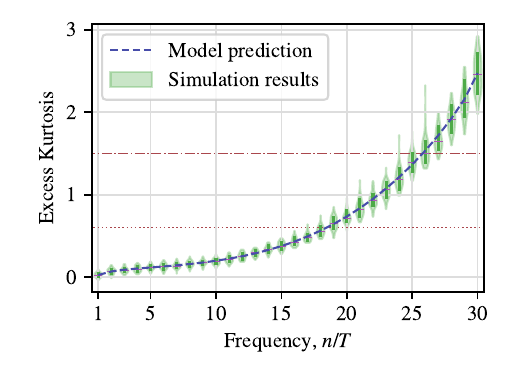}
    \caption{The distribution of excess kurtosis of all 67 pulsars by frequency. The results of our simulation match analytic predictions very closely. Color and bar conventions for the simulation results and prediction match \autoref{fig:ex-kurt-test}. To provide a reference point for excess kurtosis, the horizontal red lines indicate that value for a complex logistic (dotted) and complex Laplacian (dash-dotted) distribution.}
    \label{fig:exkurt-realistic}
\end{figure}

Similarly to what we did for the toy model of the previous section, we start by studying the excess kurtosis. Here we generate $6,000$ realizations across $30$ frequencies for all $67$ pulsars. The distribution of excess kurtosis by frequency is found in \autoref{fig:exkurt-realistic}.
The distributions are all leptokurtic (i.e., more likely to produce outliers than a Gaussian distribution), with $\bar\kappa$ growing monotonically with frequency. The total excess kurtosis is dominated by ``covariance'' terms at lower frequencies (i.e., the second term in the numerator in \autoref{eq:ex_kurt}), turning over to being dominated by Poisson-like terms (the first term in the numerator in \autoref{eq:ex_kurt}) after $f\approx 16/T$. The precise shape of the excess kurtosis curve depends on the parameters of the model; the impacts on the excess kurtosis of SMBHB parameters in more realistic models of GWs will be the subject of a future work.

To understand the origin of this deviation from Gaussianity, it is useful to rewrite the kurtosis for a complex Gaussian variable, $Z=X+iY$, as 
\begin{align}\label{eqn:kappa}
    \kappa &= \frac{3\left\langle X^2\right\rangle^2+3\left\langle Y^2\right\rangle^2+2\left\langle X^2\right\rangle\left\langle Y^2\right\rangle}{\left(\left\langle X^2\right\rangle + \left\langle Y^2\right\rangle\right)^2} + \nonumber\\&\qquad\frac{\bar\kappa_X\left\langle X^2\right\rangle^2+\bar\kappa_Y\left\langle Y^2\right\rangle^2+4\left\langle XY\right\rangle^2}{\left(\left\langle X^2\right\rangle + \left\langle Y^2\right\rangle\right)^2}\,,
\end{align}
where $\bar\kappa_X,\bar\kappa_Y$ are the excess kurtoses for $X,Y$. Consider now the case of circular $Z$. Then, $\left\langle X^2\right\rangle=\left\langle Y^2\right\rangle$ and so the first term above simplifies to $2$. The second term contains information about the excess kurtoses in $X$ and $Y$, which should be the same if $Z$ is circular, as well as the covariance between the real and imaginary parts, which should be zero if $Z$ is circular. If $Z$ is additionally Gaussian, the excess kurtoses should be zero, and we recover $\kappa=2$ as expected. Examining the above equation, we see that any of $\bar\kappa_X,\bar\kappa_Y\neq 0$, $\left<X^2\right>\neq\left<Y^2\right>$, or $\left<XY\right>\neq 0$ can lead to a non-zero excess kurtosis for the complex variable $Z$.\footnote{Note that $\left<X^2\right>\neq\left<Y^2\right>$ alone guarantees that the first term in \autoref{eqn:kappa} is greater than $2$.}

The results of our simulations show that $\bar\kappa_X,\bar\kappa_Y>0$ at all frequencies, with the $\bar\kappa_X$ typically larger, and $\left<X^2\right>\lesssim \left<Y^2\right>$ at all frequencies except the first. For all frequencies, $X$ and $Y$ are effectively uncorrelated. By the analysis of the preceding paragraph, we then expect $\bar\kappa>0$, since in the second term all contributions are either positive or zero. This is in agreement with what we see in \autoref{fig:exkurt-realistic}. Examining \autoref{eq:ex_kurt}, we can see that every term in the numerator and denominator is guaranteed to be positive, regardless of the parameters we choose for our astrophysical model. Additionally, the number of GW sources per realization is Poisson-distributed, which is always leptokurtic (i.e. has positive excess kurtosis). As a consequence, we expect $\bar\kappa>0$ generally.
    
\begin{figure}
    \centering
    \includegraphics[width=0.95\linewidth]{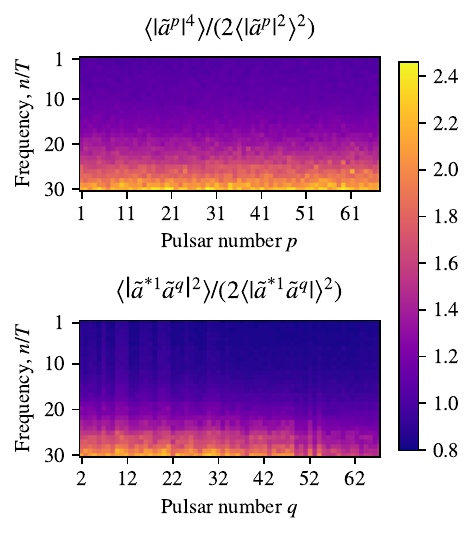}
    \caption{The ratio of fourth moments to twice the second moments for auto-pulsars (upper) and a fixed-partner subset of cross-pulsars (lower) for an astrophysical model of a background and using the 67 pulsars of the NANOGrav 15yr dataset. Pulsars \#2 through \#67 are ordered according to their angular distance on the sky from pulsar \#1.} For the auto-pulsar results, disagreement generally grows with frequency, indicating that even in the large-source limit, some non-Gaussianity persists, as discussed in \autoref{ssec:many-sources-variance}. The cross-pulsar result also shows a general growth in disagreement with frequency, but less strongly with increasing angular separation.
    \label{fig:auto_cross_realistic}
\end{figure}

Now, consider the auto- and cross-correlations of the Fourier coefficients. We find the second and fourth moments of all 67 auto-pulsar cases and the 66 cross-pulsar cases obtained by fixing one of the pulsars (without loss of generality, pulsar \#1) across $6,000$ realizations. he remaining 66 pulsars are ordered by their angular separation on the sky from this fixed pulsar, such that pulsar \#2 is closest (at $\approx 0.08\,\text{rad}$) and pulsar \#67 is farthest ($\approx 2.90\,\text{rad}$). We then calculate the ratio of fourth moments to twice the second moments and plot them in \autoref{fig:auto_cross_realistic}. If the signal is effectively Gaussian, we expect this ratio to be $1$.
The auto-pulsar results appear Gaussian at low frequencies, but the ratio grows with frequency, indicating that the fourth moment has non-Gaussian terms which become important at large frequencies. This is consistent with the discussion in \autoref{ssec:many-sources-variance}, as well as the growth in excess kurtosis with frequency seen in \autoref{fig:exkurt-realistic}. 

The cross-pulsar results display less growth in non-Gaussianity with frequency, as well as less uniform changes, with some ratios remaining close to unity across all frequencies. In general, we see more growth in non-Gaussianity when the pulsars are closer together on the sky; we note that pulsar \#50 is $\approx \pi/2$ separated from pulsar \#1, close to where the tendency for less growth begins to take over. Angular dependence of the cross-pulsar moments can be seen, e.g., in \autoref{eq:4th-moment-cross}, although how the angle enters is complicated. Further analytic investigation of the pulsar separation on sensitivity to non-Gaussianity will be the subject of a future work.

Finally, we use this realistic population to test the Cauchy distribution prediction introduced in \autoref{eq:finite-cauchy}. As a reminder, for a circular Gaussian random variable, we expect its complex argument to be uniformly distributed, and thus the tangent thereof (which is directly related to the ratio of imaginary and real parts of the Fourier coefficients) to be Cauchy-distributed with a location parameter (mean) of zero and a scale parameter of unity. From \autoref{fig:auto_cross_realistic}, we expect the distribution of $\tan\varphi$ to best match this prediction for the second through thirteenth frequencies.

\begin{figure}
    \centering
    \includegraphics[width=1.00\linewidth]{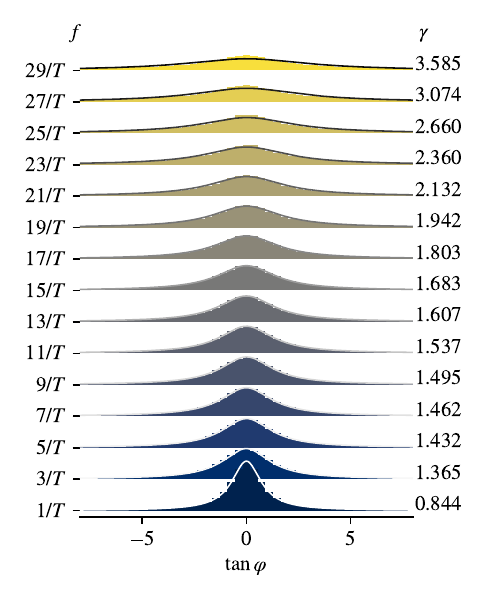}
    \caption{Histograms of the $\tan\varphi$ for all $6,000$ realizations of all $67$ pulsars by frequency, compared to the averaged best-fit Cauchy distributions by frequency. A circular Gaussian distribution in the Fourier coefficients would lead to a mean of zero and scale parameter $\gamma=1$, but mild correlations and slightly unequal variances in the realistic model's coefficients cause deviations from this idealized case, mostly in the scale parameter. We show only odd frequencies as a space-saving measure, but the trend in even frequencies is consistent with what is shown.}
    \label{fig:cauchy}
\end{figure}

To test this, we proceed as follows. We find the best fit parameters, assuming a Cauchy distribution, to the distribution of the $6,000$ realizations of $\tan\varphi_i^p$, indexing across all pulsars $p$ and frequencies $f_i$. We then combine the data across pulsars and average the best fits to obtain a mean Cauchy location and scale parameter per frequency.\footnote{The weighted sum of Cauchy-distributed independent random variables is Cauchy-distributed, with location and scale parameters given by the identically-weighted sums of the location and scale parameters of the same.} At each frequency, we then compare the data distribution to the predicted Cauchy distribution using these averaged parameters, as shown in \autoref{fig:cauchy}.

We see here similar behavior as in \autoref{fig:auto_cross_realistic}. The first frequency has location and scale parameters $\approx(1.30\cdot 10^{-4},0.844)$, fairly close to the predicted (circular Gaussian) result of $(0,1)$. As frequency increases, the location parameter (mean) stays effectively zero, while the scale parameter monotonically increases. Thus, the distribution of $\tan\theta$ is over a broader range at higher frequencies compared to lower frequencies. We can interpret the frequency-dependent increase in non-Gaussianity in the auto-pulsar results, as shown in \autoref{fig:auto_cross_realistic}, as being mainly due to unequal variances between real and imaginary parts of the Fourier coefficients. In particular, the imaginary part of the coefficients tends to have higher variance in this realization of our model (see \autoref{eqn:scale-shape_Z} and surrounding discussion).

\section{Discussion}\label{sec:conclusion}
We have developed a general framework for calculating the variance and kurtosis of a single pulsar's timing-residual Fourier coefficients in response to a GWB from a finite population of sources. We applied it to both a toy model and a more astrophysical model of an SMBHB population and compared limits of our model to a Gaussian GWB. For the first time, our work accounts for sources emitting at non-central frequencies via the explicit inclusion of window functions. Compared to previous work on the subject, our simple analytical formalism provides intuitive predictions for the level of non-Gaussianities expected to be produced by SMBHB populations.
    
We have analytically computed the excess kurtosis for the Fourier coefficients of an astrophysical GWB, and shown that for a finitely-sourced GWB, its always leptokurtic (i.e., more likely to produce outliers than a Gaussian distribution). In the large-source limit, the moments tend towards the moments of a Gaussian random variable.

Complementing the moments on the magnitudes of the Fourier coefficients, we have shown that the distribution of the tangent of their arguments should follow a standard Cauchy distribution. The distributions on magnitudes and arguments combined paves the way for quickly generating the timing-residual Fourier coefficients of a realistic SMBHB background.
    
Our work highlights the need for developments in current astrophysical inference. Firstly, we have demonstrated the importance of not assuming that all sources emit GW at harmonics of $f_n=n/T$ for $n\in\mathbb{N}^+$. This should be easy to implement with a SAM model assuming that the appropriate window function $w$ is applied. 

In PTA data analyses, the Gaussian assumption is achieved by marginalizing the Fourier coefficients in a hierarchical likelihood over a Gaussian prior with variance $S(f)\Delta f$ \citep{van2015low}. The PSD $S(f)$ is modeled as a frequency-dependent spectrum, e.g., a power-law conditioned on hyperparameters to be estimated by MCMC. To deal with non-Gaussianity, future analyses should model the GWB as a combination of a Gaussian GWB and individual continuous-wave sources \citep{becsy2020joint, becsy2022exploring}, with further development possible to model non-Gaussianities in regimes with a low number of unresolvable binaries \citep{cornish2015gravitational}.

In addition, PTA detection relies on Hellings--Downs correlations between pulsars. We observed that the angular separation between pulsars affects the sensitivity to non-Gaussianity in cross-pulsar moments, with more-separated pulsars generally showing less sensitivity. Thus a detection of non-Gaussianity would rely more heavily on pulsars nearby one another on the sky. Additional work is necessary to translate our results into statements about the impact on detectability; for related work on pulsar variance, see Sec.~VI of \cite{allen_valtolina_24}, and \cite{Kuntz:2026usl} for discussion of a generalized Hellings--Downs four-pulsar correlator and consequences for detectability.

Moving forward, analyses aimed at detecting non-Gaussian features imprinted in the GWB by the discrete nature of SMBHBs (see Ref.~\cite{falxa2025modeling} and \citep{raidal2026heavy} for examples of a search strategy) could provide a path to identifying the origin of the GWB. Indeed, in this work, we provide a clear target for the level of non-Gaussianity expected for an SMBHB-sourced GWB. Detecting these features would be a strong indication of an astrophysical origin of the GWB. On the other hand, a lack of detection could point to a primordial origin.

Additional complexities are introduced with more astrophysical models, where the amplitudes depend on physical parameters such as chirp mass and luminosity distance, and which may account for the inclination, polarization angle, or eccentricity of the source. However, the specific astrophysical model we studied here shows that our results hold even in more realistic scenarios. In future work, we will generalize the astrophysical model to explore the moments of Fourier coefficients and PSDs for eccentric binaries, which we expect to increase the level of non-Gaussianity in the GWB because of its additional degrees of freedom. Adding eccentricity effects will also introduce inter-frequency covariances and non-stationarity \citep{Reardon:2023gzh, falxa2024modeling}, which will need to be modeled with our framework.

Our work highlights the current issues in astrophysical inference of PTA data sets and introduces some potential solutions for accurate, but computationally tractable, analyses. With further development, our work can be extended to any finite population of GW sources across the GW frequency spectrum and with any GW detector, to determine the source, and to characterize the sources, of gravitational wave backgrounds.

\section*{Acknowledgements}

\textit{Discussions:} We thank Bruce Allen, Alexander Criswell, Celia Fielding, Kyle Gersbach, Jeffrey Hazboun, Thomas Konstandin, Nima Laal, Matthew Miles, Polina Petrov, Kai Schmitz, Levi Schult, and Serena Valtolina for useful conversations. We thank Kayhan G\"ultekin for early collaboration and extensive comments on a draft of this paper.

\textit{Funding/resources:} Our work was supported by the NANOGrav NSF Physics Frontier Center awards \#2020265 and \#1430284. S.R.T acknowledges support from NSF AST-2307719, and an NSF CAREER \#2146016. S.R.T is also grateful for support from a Vanderbilt University Chancellor's Faculty Fellowship. E.L.H. acknowledges summer research support from the Dean's Office of the Wentworth School of Sciences \& Humanities. S.C.S acknowledges awards from NSF \#2309246. J.M.W. acknowledges support from the APS--Simons Travel \& Professional Development Awards. This work was conducted in part using the resources of the Advanced Computing Center for Research and Education (ACCRE) at Vanderbilt University, Nashville, TN. W.G.L thanks the Fisk-Vanderbilt Bridge Program for their computing support. W.G.L. and S.R.T. acknowledge the Establishing Multimessenger astronomy Inclusive Training (EMIT) program, NSF NRT-2125764.

\appendix

\section{Ensemble averaging over realizations of finite populations}\label{app:ensemble}

To compute the moments of $\tilde{a}_i^p$, we must ensemble average over quantities that describe the source, such as the wave phase $\phi_j$, GW propagation direction $\hat\Omega_j$, inclination angle $\iota_j$, polarization angle $\psi_j$, and amplitude parameters $\vec\theta_j$, where $j$ indexes the sources. That average is taken over a probability measure $p(X)$; for a generic observable $Q$,
\begin{equation}
    \langle Q\rangle_{X} = \int p(x) Q(x) dx, \ \mathrm{ where } \, \int p(x)dx = 1.
\end{equation}
We begin by deriving the fourth moment of $\tilde{a}_{i,\mathrm{unpol}}^p$, which requires taking the expectation of $\left|\tilde{a}_{i,\mathrm{unpol}}^p\right|^4$,
\begin{widetext}
    \begin{align}\label{eq:fourth-prod}
        \left|\tilde{a}_{i,\mathrm{unpol}}^p\right|^4 &= \sum_j \frac{\mathcal{A}_j^4}{256\pi^4f_j^4}\left|R_{pj}\right|^4 \left[\left(w_j^{--} + w_j^{++}\right)^2 + 2w_j^{-+}w_j^{-+} + a\left(e^{i\phi_j}, e^{-i\phi_j}\right)\right] \nonumber \\
        &+ \sum_{j\neq k} \frac{\mathcal{A}_j^4}{256\pi^4f_j^2f_k^2} \left|R_{pj}\right|^2\left|R_{pk}\right|^2 \left[\left(w_j^{--} + w_j^{++}\right)\left(w_k^{--} + w_k^{++}\right) + b\left(e^{i\phi_j}, e^{-i\phi_j}\right) b\left(e^{i\phi_k}, e^{-i\phi_k}\right)\right] \nonumber \\
        &+ \sum_{j\neq k}  \sum_{l\neq m} \frac{\mathcal{A}_j\mathcal{A}_k\mathcal{A}_l\mathcal{A}_m}{256\pi^4f_jf_kf_lf_m}\left(w_j^-R_{pj}e^{i\phi_j} - w_j^+R^*_{pj}e^{-i\phi_j}\right) \left(w_k^-R^*_{pk}e^{-i\phi_j} - w_k^+R_{pk}e^{i\phi_k}\right) \nonumber \\
&\qquad\qquad\quad\qquad\qquad\qquad \times \left(w_l^-R^*_{pl}e^{-i\phi_l} - w_l^+R_{pl}e^{i\phi_l}\right) \left(w_m^-R_{pm}e^{i\phi_m} - w_m^+R^*_{pm}e^{-i\phi_m}\right),
    \end{align}
\end{widetext}
where $a$ and $b$ are functions with an odd number of $e^{i\phi_j}$ or $e^{i\phi_j}$ terms. These terms do not contribute to the fourth moment after averaging over wave phase $\phi\in[0, 2\pi)$. When ensemble averaging over $\phi$, we utilize the following identities,
\begin{align}\label{eqn:exp-avg}
    &\left<e^{i\phi_j}\right>_\phi = \int_0^{2\pi} e^{i\phi_j} \frac{d\phi_j}{2\pi} = 0 \\
    &\left<e^{i(\phi_{j}-\phi_k)}\right>_\phi = \iint_0^{2\pi} e^{i(\phi_{j}-\phi_k)} \frac{\mathrm{d}\phi_j}{2\pi} \frac{\mathrm{d}\phi_k}{2\pi} = \delta_{jk} \\
    &\left<e^{i(\phi_{j}-\phi_k - \phi_l + \phi_m)}\right>_\phi = \delta_{jl}\delta_{km} \text{ with } j\neq k, l\neq m\,.
\end{align}

This removes covariant phase functions and the second double summation in the final line of \autoref{eq:fourth-prod}. We get,
\begin{widetext}
    \begin{align}\label{eq:4th-moment-phaseavg}
        \left<\left|\tilde{a}_{i,\mathrm{unpol}}^p\right|^4\right>_\phi &= \sum_j \frac{\mathcal{A}_j^4}{256\pi^4f_j^4}\left|R_{pj}\right|^4 \left[\left(w_j^{--} + w_j^{++}\right)^2 + 2w_j^{-+}w_j^{-+}\right] \nonumber \\
        &+ \sum_{j\neq k} \frac{2\mathcal{A}_j^4}{256\pi^4f_j^2f_k^2} \left|R_{pj}\right|^2\left|R_{pk}\right|^2 \left(w_j^{--} + w_j^{++}\right)\left(w_k^{--} + w_k^{++}\right)  \nonumber \\
    \end{align}
\end{widetext}

Next, we average over the GW propagation direction $\hat\Omega$. We must find the following two quantities: $\left<\left|R_{pj}\right|^2\right>_{\hat{\Omega}}$ and $\left<\left|R_{pj}\right|^4\right>_{\hat{\Omega}}$. When we cross-correlate the (complex) antenna response $R_p = R_p^+-iR_p^\times$ with sources $j,k$ at pulsars $p$ and $q$, we have the following averaging:
\begin{align}\label{eq:cross_antenna}
    \left<R^*_{p,j} R_{q,k}\right>_{\hat{\Omega}} &= \left<\chi_{p,j}^*\chi_{q,k} \sum_A F_p^A(\hat{\Omega}_j) F_q^A(\hat{\Omega}_k) \right>_{\hat{\Omega}} \\
    &\approx \left<\chi_{p,j}^*\chi_{q,k}\right>_{\hat{\Omega}} \left<\sum_A F_p^A(\hat{\Omega}_j) F_q^A(\hat{\Omega}_k) \right>_{\hat{\Omega}}\,. \nonumber
\end{align}
The exponential functions oscillate rapidly as a function of $\hat{\Omega}$, while the expectation of the product of antenna responses $F_p^A$ varies relatively slowly. Hence, we can separate the expectation into two products as shown in the second line of \autoref{eq:cross_antenna}. The expectation is
\begin{align}\label{eq:avg_transfer}
    \left<\chi_{p,j}^*\chi_{q,j}\right>_{\hat{\Omega}} \equiv \left<\chi_{pq}\right>_{\hat{\Omega}} 
    &\approx
    \begin{cases}
        2 &\text{ if } p=q \\
        1 &\text{ else.}
    \end{cases}
\end{align}
This result is an approximation because the cross-terms generate cosine functions. With current PTAs of $T=\mathcal{O}(10\mathrm{yr})$ and $L_p>1\,\text{kpc}$, this is a rapidly oscillating function and can be ignored. However, in distant future data sets, this oscillation term will become relevant~\citep{mingarelli2014effect}. 

For a GWB measurement via a pulsar $p$ with antenna response functions $F_p=F^+_p-iF^\times_p$, we can define the unpolarized, polarized, and cross-polarized responses for a pulsar pair as being proportional to $\operatorname{Re}[F^*_pF_q]$, $\operatorname{Im}[F_pF_q]$, and $\operatorname{Abs}[F_pF_q]$, respectively. The moments of these responses allow us to calculate the moments of the Fourier coefficients.

For an unpolarized GWB, the mean of the product of antenna responses $\mu_u(\gamma_{pq})$ is given by
\begin{align}
    \mu_u(\gamma_{pq}) &= \frac{1}{4\pi} \int d\hat{\Omega} \sum_{A\in\{+,\times\}} F_p^A(\hat{\Omega}) F_q^A(\hat{\Omega}) \nonumber \\
    &= \frac{1}{4\pi} \int d\hat{\Omega}\, \mathrm{Re}[F_p^*(\hat\Omega)F_q( \hat\Omega)] \nonumber \\
    &= \frac{1}{4} + \frac{1}{12}\cos\gamma_{pq} + \frac{1}{2}\Psi(\gamma_{pq})\,,\\
    \Psi(\gamma_{pq}) &= \left(1-\cos\gamma_{pq}\right)\log\left(\frac{1-\cos\gamma_{pq}}{2}\right)\,.
\end{align}
This is called the Hellings-\&-Downs (HD) correlation \citep{hellingsdowns}, and it is the signature correlation of a GWB.
        
Now, \autoref{eq:cross_antenna} simplifies to \citep{allen2023variance}
\begin{align}
    \left<R^*_{pj}R_{qj}\right> = \left<\chi_{pq}\right> \mu_u(\gamma_{pq})
\end{align} 
for a single source $j$.

Finally, we choose to normalize the inter-pulsar correlation such that the auto-correlation term is equal to 1 when pulsars $p=q$, as is common in the literature. This allows us to write cross-correlations as a fraction of the autocorrelation,
\begin{align}\label{eq:normalised_hd}
    \Gamma_{pq} = \frac{3}{2}\mu_u(\gamma_{pq})
\end{align}
We now turn our attention to the ensemble average of $\left|R_{pj}\right|^4$. We start by looking at $\left<\left|R_{pj}\right|^2\left|R_{qj}\right|^2\right>_{\hat{\Omega}}$ by applying results from \citet{allen2023variance}. We can rewrite this four-point function in terms of real and imaginary components,
\begin{align}
    &\left<\left|R_{pj}\right|^2\left|R_{qj}\right|^2\right>_{\hat{\Omega}} = \left<\left|R^*_{pj}R_{qj}\right|^2\right>_{\hat{\Omega}} \\
    &\approx \left<|\chi_{pq}|^2\right>\left(\left<\mathrm{Re}\left[F^*_{pj}F_{qj}\right]^2\right>_{\hat{\Omega}} + \left<\mathrm{Im}\left[F^*_{pj}F_{qj}\right]^2\right>_{\hat{\Omega}}\right) \nonumber.
\end{align}
The real term represents the second moment of the ``unpolarized" Hellings-\&-Downs correlation, while the imaginary term represents the second moment of the ``polarized" Hellings-\&-Downs correlation. We will deal with each quantity separately. When we find higher moments of a point source, we require,
\begin{align}\label{eq:2nd-transfer}
    \left<|\chi_{pj}^*\chi_{qj}|^2\right>_{\hat{\Omega}} \equiv \left<|\chi_{pq}|^2\right>
    \approx \begin{cases}
        6 &\text{ if } p=q\\
        4 &\text{ else}.
    \end{cases}
\end{align}
In this case, the cross-terms that we ignored because they rapidly oscillated in the mean now cross with their conjugates, and contribute towards the average.

\begin{figure}
    \centering
    \includegraphics[width=0.9\linewidth]{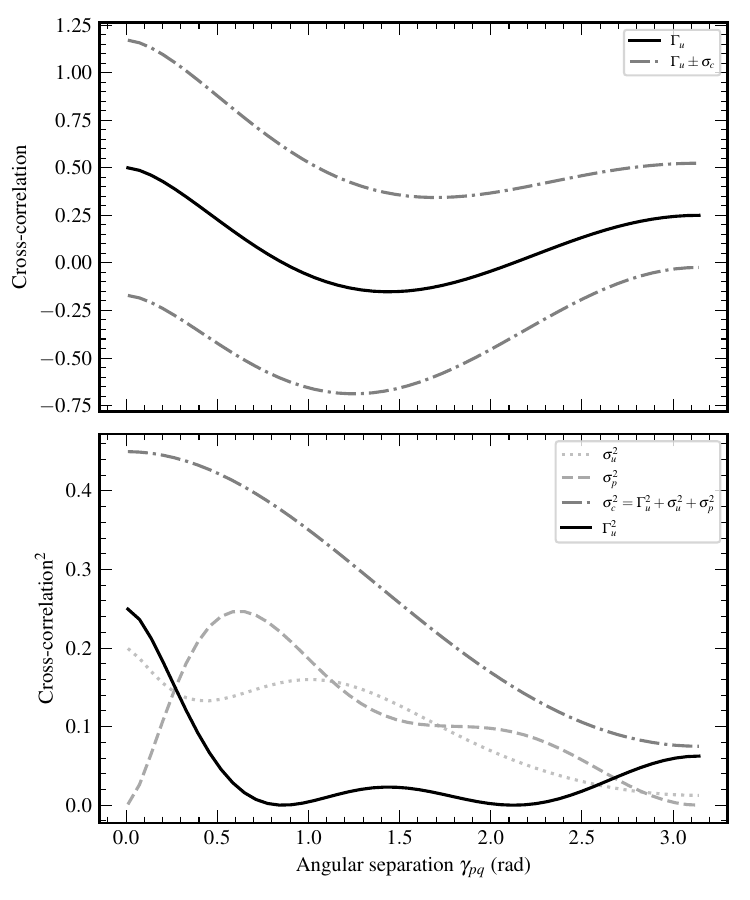}
    \caption{Top: The Hellings-\&-Downs curve as a function of angular separation between pulsars $p,q$ (solid line) plus/minus a standard deviation (dot-dashed). Bottom: A figure of $(\Gamma_{pq})^2$ (solid), $\sigma^2_u$ (dotted), $\sigma^2_p$ (dashed), and $\sigma^2_c$ (dot-dashed) as a function of angular separation $\gamma_{pq}$.}\label{fig:HD}
\end{figure}

The second moment of the unpolarized Hellings-\&-Downs correlation is,
\begin{align}
    \left<\mathrm{Re}\left[F^*_{pj}F_{qj}\right]^2\right>_{\hat{\Omega}} = \mu_u^2(\gamma_{pq}) + \sigma^2_u(\gamma_{pq}),
\end{align}
where $\sigma_u^2(\gamma_{pq})$ is the variance of the unpolarized Hellings-\&-Downs correlation, which is derived in Appendix D of \citet{allen2023variance}. We give its form here; however, to be consistent with the normalization of \autoref{eq:normalised_hd}, we also normalize the variance.
\begin{align}\label{eq:var_hd}
    \left(\frac{3}{2}\right)^2\sigma_{u}^2(\gamma_{pq}) &= \left(\frac{873}{320}+\frac{3}{32}\cos(\gamma_{pq}) - \frac{839}{320}\cos^2\gamma_{pq}\right) \nonumber \\
    & + \frac{3}{16}\left(18 - 10\cos\gamma_{pq} -3\Psi(\gamma_{pq})\right)\Psi(\gamma_{pq})
\end{align}
Similarly, the second moment of the polarized Hellings-\&-Downs correlation is,
\begin{align}
    \left<\mathrm{Im}\left[F^*_{pj}F_{qj}\right]^2\right>_{\hat{\Omega}} = \mu_p^2(\gamma_{pq}) + \sigma^2_p(\gamma_{pq}).
\end{align}
The mean of polarized sources $\mu_p(\gamma_{pq})=0\,\forall\,\gamma_{pq}$, which can be understood by example: one origin of polarization might be the inclination of the binary's orbital plane with respect to the observation line, but assuming a uniformly-distributed inclination angle, on average the plane is not inclined and thus the wave is not polarized. The variance $\sigma_p^2(\gamma_{pq})$ of the polarized Hellings-\&-Downs correlation is derived in Appendix E of \citet{allen2023variance}. For consistency, we also normalize this quantity,
\begin{align}\label{eq:var_polarised}
    \left(\frac{3}{2}\right)^2\sigma^2_{p}(\gamma_{pq}) &=  \frac{21}{8} \left(\cos^2\gamma_{pq} - 1\right) \nonumber \\
    &+ \frac{9}{16} \left(3\cos\gamma_{pq}-7\right)\Psi(\gamma_{pq}).
\end{align}
Bringing everything together, we find that,
\begin{align}\label{eq:cross-pol}
    \left<|R_{pj}|^2|R_{qj}|^2\right>_{\Omega} &= \left<|\chi_{pq}|^2\right>\left(\Gamma_{pq}^2 + \sigma_u^2(\gamma_{pq}) + \sigma_p^2(\gamma_{pq}) \right) \nonumber \\
    &= \left(\frac{3}{2}\right)^2\left<|\chi_{pq}|^2\right> \sigma_c^2(\gamma_{pq}),
\end{align}
where $\sigma_c^2(\gamma_{pq})$ is the variance of the cross-polarized Hellings-\&-Downs correlation, which is derived in Appendix F of \citet{allen2023variance}. We have shown that this quantity is the sum of the variances of the real and imaginary parts of the cross-correlation of the antenna response functions.
\begin{equation}\label{eq:variance-cross-pol}
   \left(\frac{3}{2}\right)^2 \sigma_c^2(\gamma_{pq}) = \frac{39}{160}+ \frac{3}{16}\cos\gamma_{pq}  + \frac{3}{160} \cos^2\gamma_{pq}\,.
\end{equation}
We plot these various quantities as a function of $\gamma_{pq}$ in \autoref{fig:HD}. Finally, to ensure that we have normalized our quantities correctly, we must substitute
\begin{equation}
    \left<R^*_{pj} R_{qj}\right> \to \frac{2}{3} \left<\frac{3}{2} R^*_{pj} R_{qj}\right>,
\end{equation}
in any ensemble average that we compute.

Therefore, substituting these quantities into \autoref{eq:4th-moment-phaseavg} and setting $p=q$ (i.e., $\gamma_{pq}=0$),
\begin{widetext}\begin{equation}
    \left<|\tilde{a}_{i,\mathrm{unpol}}^p|^4\right>_{\phi,\hat\Omega} = 2\left<|\tilde{a}_{i,\mathrm{unpol}}^p|^2\right>_{\phi,\hat\Omega}^2 + \frac{7}{10}\sum_{j} \frac{\mathcal{A}_j^4}{24^2\pi^4 f_j^4} \left[(w_{j}^{--} + w_{j}^{++})^2 + 2w_{j}^{--}w_{j}^{++}\right] + 4\left(\sum_{j} \frac{\mathcal{A}_j^2}{24\pi^2 f^2_j}w_{j}^{-+}\right)^2.  \label{eq:4th-moment}
\end{equation}\end{widetext}

Finally, we detail how we find the coefficient of the second term (which we label as the `Poisson term') in \autoref{eq:4th-poisson-pol}. Following the derivation found in \citet{allen_valtolina_24}, we replace the factor of 0.7 in \autoref{eq:4th-moment} with the following ensemble average:
\begin{equation}
    \left<\left|\frac{1}{16i}\left(1+\cos\iota_j\right)^2\,e^{2i\phi_j} F_j + \mathrm{c.c.}\right|^4|\chi_{p}|^4\right> \equiv \left<P\right>.
\end{equation}
We start by expanding the fourth power and averaging over $\psi_j$. Terms proportional to $e^{\pm4i\psi_j}$ are eliminated, therefore, we find,
\begin{align}
    \left<P\right>_{\psi} = &
\frac{1}{16^4}\left(\left(1+x\right)^8+\left(1-x\right)^8 \nonumber \right. \\
    &+ \left. 4\left(1+x\right)^4\left(1-x\right)^4\right)|\chi_{pj} F_{pj}|^4,
\end{align}
where we have substituted $x=\cos\iota$. We now ensemble average over $\iota$ using the following integrals:
\begin{subequations}
\begin{align}
    \left<\left(1\pm x\right)^8\right>_\iota &= \frac{256}{9}\, \\
    \left<\left(1+x\right)^4\left(1-x\right)^4\right>_\iota &= \frac{128}{315}.
\end{align}
\end{subequations}
Additionally, $\chi_{pj} F_{pj} = R_{pj}$. Combining the above relations with \autoref{eq:2nd-transfer} ,  we find,
\begin{align}
    \left<P\right> = \left<P\right>_{\psi,\iota,\hat{\Omega}} &= \frac{6}{16^4}\left(\frac{512}{9}+4\frac{128}{315}\right)\sigma_c^2(0) \nonumber \\
    &= \frac{1}{24^2} \frac{108}{175}
\end{align}
Hence,
\begin{widetext}\begin{equation}
    \left<|\tilde{a}_i^p|^4\right>_{\phi,\hat\Omega,\psi,\iota} = 2\left<|\tilde{a}_i^p|^2\right>_{\phi,\hat\Omega,\psi,\iota}^2 + \frac{108}{175}\sum_{j} \frac{A_j^4}{24^2\pi^4 f_j^4} \left[(w_{j}^{--} + w_{j}^{++})^2 + 2w_{j}^{--}w_{j}^{++}\right] + 4\left(\sum_{j} \frac{A_j^2}{24\pi^2 f^2_j}w_{j}^{-+}\right)^2.  \label{eq:4th-moment-pol-N}
\end{equation}\end{widetext}

The final step finds the ensemble average over the number of binaries $N$ in each realization. We detail this process in \autoref{app:sam}, and the final fourth moment is given by \autoref{eq:4th-poisson-pol}.

By way of example of the utility of these relationships, we present below the fourth cross-pulsar moment of the Fourier coefficients $\tilde{a}_{i,\mathrm{unpol}}^p$:
\begin{widetext}\begin{align}
    \left<|\tilde{a}_{i,\mathrm{unpol}}^{*p}\tilde{a}_{i,\mathrm{unpol}}^q|^2\right>_{\phi,\hat\Omega} &= \sum_{j} \frac{\mathcal{A}_j^4}{24^2\pi^4 f^4} \left[\left<|\chi_{pp}|^2\right>\sigma^2_c(\gamma_{pq})  \left(w_{j}^{--} + w_{j}^{++}\right)^2 + \left<|\chi_{pq}|^2\right>(4\Gamma_{pq}^2 + 2\left(2\sigma^2_u(\gamma_{pq}) - \sigma^2_c(\gamma_{pq}))\right) w_{j}^{--}w_{j}^{++} \right]\nonumber \\
    &+ \sum_{j\neq k} \frac{\mathcal{A}_j^2\mathcal{A}_k^2}{24^2\pi^4 f^2_j f_k^2} \left((1+\Gamma_{pq}^2)\left(w^{--}_{j} + w^{++}_{j}\right)\left(w^{--}_{k} + w^{++}_{k}\right) + \left(\Gamma_{pq}^2 - 1\right) \left(w^-_{j}w^+_{k} + w^+_{j}w^-_{k}\right)^2 \right)\label{eq:4th-moment-cross}
\end{align}\end{widetext}
In the limit $p=q$, this reproduces \autoref{eq:4th-moment}. We leave the general case for $\tilde{a}_i^p$ as future work.
\section{Semi-Analytic Models}\label{app:sam}

There are two quantities in our model that are dependent on astrophysical parameters $\vec{\theta}=[\mathcal{M},z,...]$: the GW strain amplitude $A(f, \vec{\theta})\equiv A$, and the \textit{mean} number of GW sources $\bar{N}(f, \hat{\Omega}, \vec{\theta})$ that emit within a frequency bin $\Delta f$, bin of GW propagation direction $\Delta\hat\Omega$, and astrophysical parameter bins $\Delta \vec{\theta}$. In this case, we can replace the summation over $N$ sources with a sum over grid cells $J$ containing a mean number of sources $\bar{N}_J$ using \autoref{eq:sum_pver_bins}, which we remind the reader below.
\begin{equation}
    \sum_{j=1}^N \frac{A_j^2}{f_j^2} \quad\Rightarrow\quad\sum_J^{n} N_J \frac{A_J^2}{f_J^2}\,.
\end{equation}
There are $n$ number of grid cells in our parameter space. In the limit $n\to\infty$, the grid cell $J$ is infinitessimally small. Hence, our Riemann sum becomes an integral. For some function $Q$,
\begin{equation}
    \lim_{n\to\infty}\sum_J^{n}N_J\, Q(f_J,\hat\Omega_J, \vec{\theta}_J) =\int \frac{dN}{df d\hat\Omega d\vec{\theta}} Q(f, \hat{\Omega}, \vec{\theta}) d\hat\Omega d\vec\theta.
\end{equation}
The infinitessimal number density $d\bar{N}/df d\hat\Omega d\vec{\theta}$ models the distribution of the number of GW sources as a function of frequency and strain amplitude parameters, which becomes our probability measure that we use to average over $\vec{\theta}$ and $\hat\Omega$. In PTA data analyses, this quantity is often described as $d\bar{N}/df_r d\hat\Omega d\log\mathcal{M}dz$, where $\mathcal{M}$ is the chirp mass of an SMBHB and $z$ is the redshift to the SMBHB \citep{Sesana:2008mz, middleton2015astrophysical, sato2024exploring}, and evaluated at the rest-frame GW frequency $f_r=f(1+z)$.

Note that in the limit $n\to\infty$, grid cell $J$ is infinitessimally small. As each GW source has a unique frequency, GW propagation direction, and amplitude, there are either $N=0$ or $N=1$ sources emitting in cell $J$ Therefore, the square of the number of sources $N^2=0$ or $N^2=1$. This is the definition of shot noise, also known as Poisson variance \citep{sato2024exploring, lamb_taylor_24}. The number of GW sources $N$ in a realization of a GWB is Poisson-distributed with mean $\bar{N}$ and it is conditioned on model hyperparameters $\vec\eta$. The hyperparameters $\vec\eta$ include parameters that limit the number of binaries above certain chirp masses and redshifts, hence, changing the distribution of sources along each astrophysical parameter, modifying the number of binaries and their amplitudes. Therefore, the mean number density of GW sources per unit frequency, GW propagation unit angle, and per astrophysical parameter is $d\bar{N}(f, \hat{\Omega},\vec{\theta}|\eta)/df d\hat\Omega d\vec{\theta}$ is simply the \textit{ensemble averaged number density} of sources for a given population model and hyperparameters $\eta$. \citep{sesana2009gravitational,sato2024exploring}. This is the semi-analytical model (SAM) of the astrophysical GWB.

To compute the ensemble average of quantities over $N$, we utilize the fact that the number of sources $N_J$ in cell $J$ is Poisson distributed with mean $\bar{N}_J$. For terms involving single summations, this process is simple -- we replace $N_J$ by $\bar{N}_J$. For example, for an arbitrary quantity $X = \sum_J N_J\,Q_J$ for some function $Q_J$ at grid cell $J$, the ensemble average of $X$ over $N$ sources is,
\begin{equation}
    \left<X\right>_N = \sum_J \bar{N}_J\,Q_J.
\end{equation}
Averaging over quantities involving double summations are a little more involved. We require finding the ensemble average of the quantity $X^2$, where
\begin{align}
    \left<X^2\right>_N = \sum_{JK} \left<N_J N_K\right> Q_J Q_K,
\end{align}
where $J, K$ are dummy variables. We note that if $J\neq K$, $\left<N_J N_K\right>=\left<N_J\right>\left<N_K\right>=\bar{N}_J\bar{N}_K$. Hence, separating the double sum into diagonal and non-diagonal components,
\begin{align}
    \left<X^2\right>_N = \sum_{J} \left<N_J^2\right> Q_J^2 + \sum_{J\neq K} \left<N_J\right>\left<N_K\right> Q_J Q_K.
\end{align}
We can then re-write the second term as the difference between a double sum and a diagonal term. Collecting similar terms, we find,
\begin{align}\label{eq:poisson_definition}
    \left<X^2\right>_N &= \sum_{JK} \left<N_J\right>\left<N_K\right> Q_J Q_K + \sum_{J} \left(\left<N_J^2\right> - \left<N_J\right>^2\right)\, Q_J^2 \nonumber \\
    &= \left<X\right>_N^2 + \sum_J \bar{N}_J Q_J,
\end{align}
where the second term on the second line results from the definition of the variance of a Poisson-distributed random variable, $\bar{N}_J = \mathrm{Var}[N_J] = \left<N_J^2\right> - \left<N_J\right>^2$.

\section{Complex random variables}\label{app:complex-rvs}

When we say that the $\tilde{a}^p_i$ are Gaussian, we mean that they follow a \emph{circular complex Gaussian distribution}. ``Complex'' here means that the random variable is complex; circularity is a property of complex random variables which says (loosely) that the statistics are invariant under arbitrary rotation. Formally, for variable $Z$ and a fixed but arbitrary $\phi\in[-\pi,\pi)$,
\begin{subequations}\begin{align}
    \left\langle Z\right\rangle &= \left\langle e^{i\phi}Z\right\rangle = e^{i\phi}\left\langle Z\right\rangle\,,\\
    \left\langle ZZ\right\rangle &= \left\langle e^{i2\phi}ZZ\right\rangle = e^{i2\phi}\left\langle ZZ\right\rangle\,.
\end{align}\end{subequations}
Because $\phi$ is arbitrary, it must be that $\left\langle Z\right\rangle=0$ and $\left\langle ZZ\right\rangle=0$. The first quantity is properly the mean, but $\left\langle ZZ\right\rangle$ is called the \emph{pseudovariance}, and has no direct analogy for real random variables. Therefore, a circular complex Gaussian has zero mean and zero pseudocovariance matrix.

The definitions of statistics for complex random variables require a more careful treatment. For a complex random variable $Z$, $\left\langle Z\right\rangle$ and $\left\langle \left(Z^*-\left\langle Z^*\right\rangle\right)\left(Z-\left\langle Z\right\rangle\right)\right\rangle$ are agreed upon as the mean and variance, respectively (and note that $\left\langle Z^*\right\rangle=\left\langle Z\right\rangle^*$). Similarly, with a second complex random variable $W$, $\left\langle \left(Z-\left\langle Z\right\rangle\right)\left(W^*-\left\langle W^*\right\rangle\right)\right\rangle$ is the covariance (and so note $\operatorname{Cov}(Z,W)=\operatorname{Cov}(W,Z)^*$). Skewness and kurtosis, which we are interested in studying, do not have unambiguous extensions. Consider, for example, $\left\langle \left|Z^*Z\right|^3\right\rangle$ as a skewness measure: since this discards phase information, it cannot determine the ``direction of the tilt'' of the distribution in the complex plane; moreover, it would return a non-zero value for a circular complex Gaussian distribution, which manifestly should have zero skewness.

To compute the (excess) kurtosis of the timing residual Fourier components, we extend the idea of the kurtosis as a ratio between a fourth central moment and the variance squared and write
\begin{equation}\label{eq:kurtosis-z_general}
    \kappa = \frac{\left\langle\left(Z^*-\left\langle Z^*\right\rangle\right)^2\left(Z-\left\langle Z\right\rangle\right)^2\right\rangle}{\sigma^4}\,,
\end{equation}
where $\sigma^2$ is the variance as defined above. This measure is real-valued and thus preserves the notion of kurtosis as a measure of tailedness of the distribution.

In the main text, we did not analytically calculate the odd moments of the $\tilde{a}^p_i$ such as skewness, which should be zero. When we perform checks of simulated $\tilde{a}^p_i$, we will use
\begin{equation}\label{eq:skewness}
    \gamma_1 = \frac{{\left\langle \left(Z-\left\langle Z\right\rangle\right)^3\right\rangle}}{\sigma^{3/2}}\,.
\end{equation}
This measure keeps full phase information and will properly return zero for a circular complex Gaussian.

As a final note, Isserlis' theorem~\citep{isserlis} works for complex variables just as well as for real variables. We can use it to show that the kurtosis of a circular complex Gaussian is \emph{not} $3$, as it is for a real univariate Gaussian. Rather, with
\begin{align}\label{eq:isserlis}
    \left\langle ZZZ^*Z^*\right\rangle &= \left\langle ZZ\right\rangle \left\langle Z^*Z^*\right\rangle+2\left\langle Z^*Z\right\rangle
\end{align}
and recalling $\left\langle Z\right\rangle=0$, $\left\langle ZZ\right\rangle=0$ if $Z$ is circular, then $\kappa=\left\langle ZZZ^*Z^*\right\rangle/\left\langle Z^*Z\right\rangle=2$. Thus, the excess kurtosis for a complex random variable is the kurtosis minus two.

The above moments can be thought of as moments of the magnitude $|Z|$ of the complex number $Z$. However, complex numbers are uniquely defined by their magnitude and their argument $\varphi=\mathrm{arg}\, Z$.\footnote{This is also called the \textit{phase} of the complex number, but to avoid confusion with the wave phase of a GW source, we will use the term \textit{argument}.}  The tangent of the argument is given by,
\begin{equation}\label{eq:complex_arg}
    \tan\varphi = \frac{\mathrm{Im}[Z]}{\mathrm{Re}[Z]}.
\end{equation}
By analyzing the distribution of the argument, we can gather more information about the distribution of the complex number, as discussed further in \aref{app:cauchy}.

\section{Combinations of window functions}\label{ssec:window-combo}

The product of coefficients gives rise to a product of window functions. For compactness, we adopt a further reduced notation,
\begin{equation}
    w^{\pm\pm} = w^\pm w^\pm = w(f\pm f_i)w(f\pm f_i)\,.
\end{equation}
In this work, we also take products of products of window functions. In this case, we are always free to redistribute the four signs as we see fit (e.g., $w^{+-}w^{+-}=w^{--}w^{++}$; note that $w^{+-}=w^{-+}$ follows from the definition). Assuming that $w^{\pm}\to\delta(f\pm f_i)$ as $T\to\infty$, then `crossed' window terms will go to zero. This is because (for $f_j>0$),
\begin{subequations}\label{eq:window_combos}\begin{align}
    w^{-+} &\to \delta(f-f_i)\delta(f+f_i) = 0\,, \\
    w^{--}w^{++} &\to \delta(f-f_i)^2\delta(f+f_i)^2=0\,.
\end{align}\end{subequations}
        
We will also often consider the \emph{one-sided} Fourier transform, which changes our range of integration in frequency from $\mathbb{R}$ to $\mathbb{R}_{\geq 0}$. Under the assumption that the window function is symmetric around $f_i=0$, the one-sided power spectrum is simply twice the two-sided power spectrum. Similarly, the one-sided fourth moment of $\tilde{a}_i^p$ is four times the two-sided quantity.

Next, consider $w^{+-}$ for various different observed GWB frequency $f$ and sample frequency $f_i$. When $f\approx f_i$, $w^+\ll w^-$ and $w^{+-}\ll w^{--}$. When $f$ and $f_i$ are well-separated (but positive), $w^+$ and $w^-$ are of comparable (small) magnitude. When computing the variance, we will often be able to neglect terms like $w^{+-}$ in comparison to terms like $w^{--}$, sometimes independently of their precise subscripts. This is the assumption often made in SAMs.

\begin{figure}
    \centering
    \includegraphics[width=\linewidth]{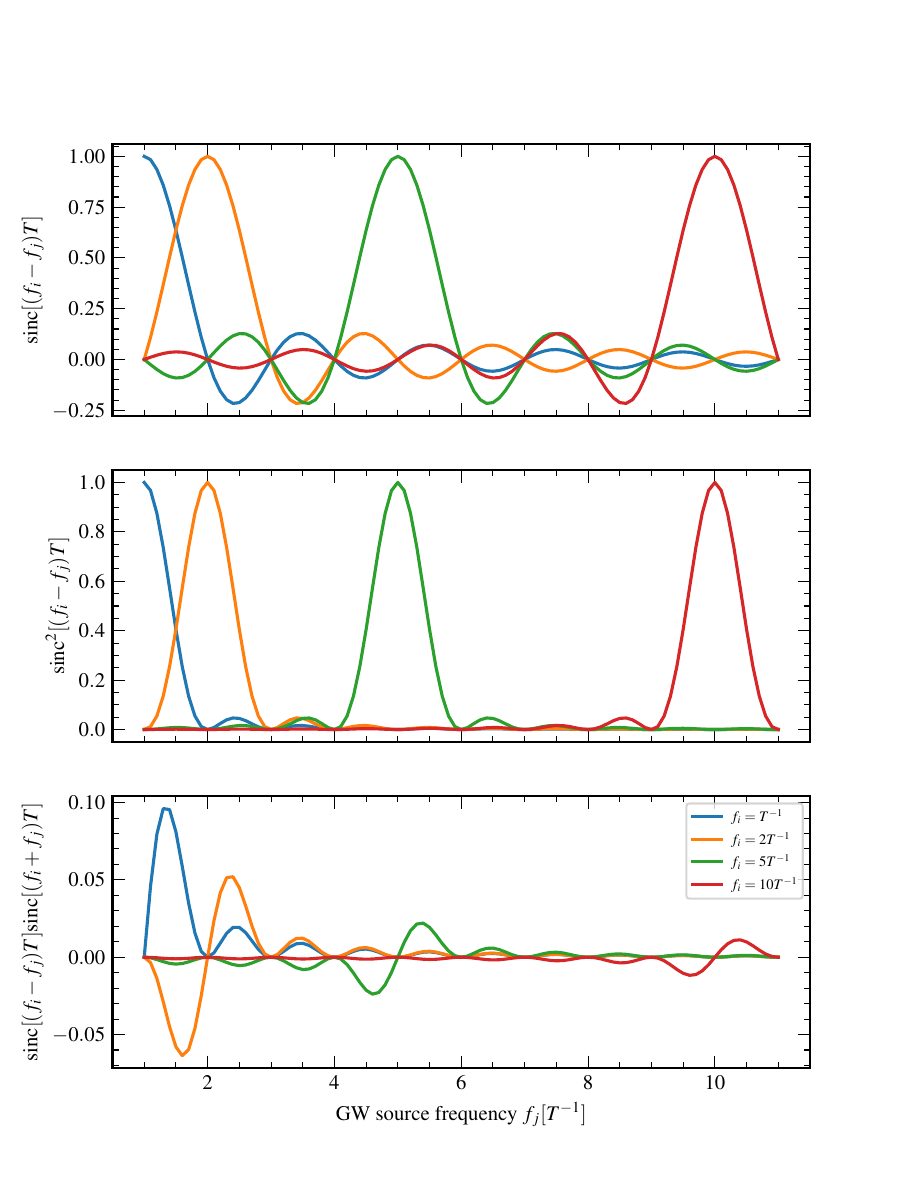}
    \caption{An illustration of $w^-$ (top), $w^{--}$ (middle), and $w^{+-}$ (bottom) for various measured frequencies as a function of GW source frequency. The first two functions strongly select at the measured frequency, but are subject to leakage from nearby frequencies; the ``crossed'' function suppresses at all frequencies but less strongly near the measured frequency.}\label{fig:sinc}
\end{figure}

Finally, when we are discussing the finite population model, we modify our notation to include the index $j$, $w_j^\pm \equiv w(f_j\pm f_i)$. \autoref{fig:sinc} shows how different combinations of $\mathrm{sinc}$ functions compare as a function of GW source frequency.

\section{Cauchy random variables}\label{app:cauchy}

This appendix serves as a brief introduction to the most relevant properties of Cauchy distributions. Most of these definitions and results can be found in a standard probability text, but are presented here for convenience.

The Cauchy distribution has two parameters: the location parameter ($x_0\in\mathbb{R}$) and the scale parameter ($\gamma\in\mathbb{R}^+$), and PDF
\begin{equation}
    f(x;x_0,\gamma) = \frac{1}{\pi\gamma}\frac{1}{1+\left(\frac{x-x_0}{\gamma}\right)^2}\,.
\end{equation}
It is sometimes called the \emph{Breit--Wigner} distribution in high-energy particle physics. We will denote a Cauchy-distributed random variable $X$ by $X\sim\operatorname{Cy}(x_0,\gamma)$. Its mean, variance, and all higher moments are undefined, but its median and mode are both simply equal to $x_0$. A last general property to note is that the (weighted) sum of Cauchy random variables is itself a Cauchy random variable: with $X_i\sim\operatorname{Cy}(x_{0,i},\gamma_i)$ for $i=1\ldots n$ and all $n$ $X$s independent,
\begin{align}
    \bar X &\equiv \sum_{i=1}^n\beta_iX_i\sim\operatorname{Cy}(\bar{x}_0,\bar\gamma) \\ 
    &\text{where}\quad \bar x_0=\sum_{i=1}^n\beta_ix_{0,i}\,,\; \bar \gamma=\sum_{i=1}^n\beta_i\gamma_i\,. \nonumber
\end{align}
Importantly, this means that the central limit theorem applied to Cauchy random variables predicts that the sum is still a Cauchy random variable, and not Gaussian. This is related to the Cauchy distribution being a stable distribution, and helps explain why we expect $\tan\varphi^p_i$ to still be Cauchy-distributed, even for $N>1$.

A construction of a Cauchy random variable, particularly relevant to this work, is as follows. Take two independent standard normal random variables: $X,Y\sim\mathcal{N}(0,1)$. Then, their ratio follows a standard Cauchy distribution: $W\equiv Y/X\sim\operatorname{Cy}(0,1)$. This can be extended to \emph{correlated} normal random variables with unequal variance as follows. First, note that if $T$ is a Cauchy random variable, then so too is $W\equiv (aT+b)/(cT+d)$ where $a,b,c,d\in\mathbb{R}$. Specifically, defining a complex Cauchy parameter $\psi=x_0+i\gamma$ and saying $T\sim\operatorname{Cy}(\psi_t)$, then $W\sim\operatorname{Cy}(\psi_w)$ where $\psi_w=(a\psi_t+b)/(c\psi_t+d)$.

Now, let $U,V$ be independent standard normal random variables and thus $T\equiv U/V$ is standard Cauchy, with $\psi_t=i$. Then $W$, by the above construction, is also Cauchy, with
\begin{align}
    \psi_w &=\frac{ac+bd+i(ad-bc)}{c^2+d^2} \\
    \rightarrow\quad x_{0w}&=\frac{ac+bd}{c^2+d^2}\,,\;\gamma_w=\frac{ad-bc}{c^2+d^2}\,.
\end{align}
We can rearrange the definition of $W$ in terms of $T$ to be
\begin{equation}
    W\equiv \frac{aU+bV}{cU+dV}=\frac{Y}{X}\,,
\end{equation}
where $Y\sim\mathcal{N}(0,\sigma_y^2=a^2+b^2)$, $X\sim\mathcal{N}(0,\sigma_x^2=c^2+d^2)$ are normal random variables with $\operatorname{Cov}[X,Y]=ac+bd$. Then noting that the correlation of $X$ and $Y$ is defined as $\rho=\operatorname{Cov}[X,Y]/\sigma_x\sigma_y$, we can write
\begin{equation}\label{eqn:scale-shape_Z}
    x_{0w}=\frac{\sigma_y}{\sigma_x}\rho\,,\;\gamma_w=\frac{\sigma_y}{\sigma_x}\sqrt{1-\rho^2}\,.
\end{equation}
Thus, with a complex Gaussian random variable $Z=X+iY$ and $\tan\varphi^p_i=\mathrm{Im}[Z]/\mathrm{Re}[Z]\leftrightarrow W=Y/X$, a nonzero location parameter indicates nonzero correlations between the real and imaginary parts of the Fourier coefficients, as seen in \autoref{fig:cauchy}. When the location parameter is (effectively) zero but the scale parameter is non-unity, this indicates that the real and imaginary parts have unequal variance. But if $Z$ is a circular complex Gaussian, then its argument will follow a standard Cauchy distribution. Deviations from standard Cauchy are signs of non-circularity, as discussed in \autoref{ssec:astro-model}.

\bibliography{apssamp}

\end{document}